\documentclass{aastex702}
\usepackage{amsmath}
\newcommand{\NGaia}{321{,}417}
\newcommand{\NFinite}{299{,}746}
\newcommand{\NMS}{218{,}654}
\newcommand{\NMSdec}{212{,}057}
\newcommand{\NSample}{210{,}191}
\newcommand{\NComp}{3{,}381{,}277}
\newcommand{\NMainSample}{1{,}880{,}195}
\newcommand{\NUsedComp}{2{,}235{,}687}
\newcommand{\NSubtiles}{35{,}285}
\newcommand{\NRealOne}{54}
\newcommand{\NOffOne}{3.4}
\newcommand{\EpochMin}{2017.7}
\newcommand{\EpochMax}{2019.5}
\newcommand{\ErrScaleK}{1.50}
\newcommand{\EcorrMed}{0.06}
\newcommand{\EOnePairs}{242{,}942}
\newcommand{\EOneDNsouth}{0.42}
\newcommand{\EOneDNnorth}{0.13}
\newcommand{\EOneNsub}{9{,}199}
\newcommand{\EOneNtile}{25{,}751}
\newcommand{\EOneNdec}{335}
\newcommand{\SigSys}{0.04}
\newcommand{\NTrue}{50}
\newcommand{\NSecure}{51}
\newcommand{\NRelNine}{48}
\newcommand{\NFalseSecure}{1.5}
\newcommand{\NCand}{75}
\newcommand{\NRxs}{55}
\newcommand{\NEra}{27}
\newcommand{\NCandNonSec}{24}
\newcommand{\NonSecSepMed}{1.8}
\newcommand{\SecSepMed}{0.20}
\newcommand{\NNonSecRxs}{8}
\newcommand{\NCandChance}{18}

\newcommand{\NCEm}{1}
\newcommand{\NCFlare}{9}
\newcommand{\NCRS}{6}
\newcommand{\NCBY}{6}
\newcommand{\NCYoung}{4}
\newcommand{\NCCV}{2}
\newcommand{\NCXA}{21}

\newcommand{\NGrpMult}{39}
\newcommand{\NGrpSingle}{4}
\newcommand{\NGrpOther}{7}

\newcommand{\NSRSC}{30}
\newcommand{\NSRSCV}{11}
\newcommand{\NDeOverlap}{41}
\newcommand{\NDeTotal}{605}
\newcommand{\NDeInSample}{66}
\newcommand{\NDeEpTwoOnly}{13}
\newcommand{\NVariable}{44}
\newcommand{\NComoving}{21}
\newcommand{\NInconclusive}{17}
\newcommand{\NSingleDet}{12}

\newcommand{\NSRSCinSample}{213}
\newcommand{\NSRSCdet}{30}
\newcommand{\NSRSCundet}{183}
\newcommand{\NSRSClow}{161}
\newcommand{\NSRSCvlass}{22}
\newcommand{\NSRSCvlassCand}{18}
\newcommand{\NSRSCcand}{45}
\newcommand{\NMultiImg}{177}
\newcommand{\NMultiDetImg}{120}
\newcommand{\NImgPerStarMin}{3}
\newcommand{\NImgPerStarMax}{7}
\newcommand{\FluxScaleMin}{1.00}
\newcommand{\FluxScaleMax}{1.05}
\newcommand{\FsAgree}{7}
\newcommand{\GNineOff}{1.4}
\newcommand{\CDOff}{2.4}
\newcommand{\LeverMin}{0.4}
\newcommand{\LeverMax}{6.8}
\newcommand{\LeverMed}{1.5}
\newcommand{\NSecureX}{49}
\newcommand{\GBmed}{14.37}
\newcommand{\GBmin}{12.86}
\newcommand{\GBmax}{15.28}

\newcommand{\NXAMult}{19}
\newcommand{\KEOne}{0}
\newcommand{\MuEOne}{3.00}
\newcommand{\EmSecName}{2MASS~J17022950+1341015}
\newcommand{\EmSecDist}{30.5}
\newcommand{\EmSecS}{1.35}
\newcommand{\EmSecRel}{0.77}
\newcommand{\EmSecRUWE}{5.3}
\newcommand{\EmSecdMG}{0.60}
\newcommand{\EmSecSNRone}{4.9}
\newcommand{\EmSecSNRlater}{1.8}
\newcommand{\NNatural}{50}

\newcommand{\SlimMed}{1.09}

\newcommand{\NConfused}{32}
\newcommand{\NDetected}{51}

\newcommand{\BarnardL}{5.1\times10^{5}}

\newcommand{\FpopFloor}{1.4\times10^{-5}}

\newcommand{\LspmChi}{77.9}
\newcommand{\LspmMu}{68}
\newcommand{\LspmMuErr}{50}
\newcommand{\FrameTieE}{+0.03}
\newcommand{\FrameTieN}{+0.01}
\newcommand{\LspmOffThree}{1.1}
\newcommand{\LspmSigThree}{7}
\newcommand{\PossIO}{20.4}
\newcommand{\PossIE}{19.9}
\newcommand{\PossIIB}{21.1}
\newcommand{\PossIIR}{20.9}
\newcommand{\LspmOffPossI}{9.9}

\newcommand{\LspmOffPS}{0.9}
\newcommand{\GaiaErrStar}{2.4}
\newcommand{\GaiaErrComp}{33}
\newcommand{\LspmChiTwoD}{104}
\newcommand{\LspmDofTwoD}{6}
\newcommand{\LspmChiRAstat}{18}
\newcommand{\LspmChiRAcomove}{16}
\newcommand{\LspmChiDecStat}{1.8}

\newcommand{\LspmQLSEdE}{0.55}
\newcommand{\LspmQLSEdN}{0.31}
\newcommand{\LspmPA}{109}
\newcommand{\LspmL}{2.1\times10^{9}}
\newcommand{\LspmOffAU}{75}
\newcommand{\LspmOffArcsec}{0.8}
\newcommand{\LspmS}{1.81}
\newcommand{\LspmSep}{0.26}
\newcommand{\LspmRel}{0.987}
\newcommand{\LspmDist}{98}
\newcommand{\LspmMaj}{3.3}
\newcommand{\LspmMin}{1.3}
\newcommand{\StackMSL}{1.2\times10^{6}}
\newcommand{\StackMSLerr}{1.9\times10^{5}}
\newcommand{\StackMSLsnr}{6.3}

\newcommand{\TRfloor}{1.2\times10^{-5}}

\newcommand{\VlassFiftyS}{0.54}
\newcommand{\VlassFiftyEIRP}{3.2\times10^{17}}
\newcommand{\BLratio}{4.6\times10^{4}}
\newcommand{\BLratioPub}{1.5\times10^{5}}
\newcommand{\BLEIRPcorr}{6.9\times10^{12}}
\newcommand{\BLEIRPpub}{2.1\times10^{12}}
\newcommand{\BLSJy}{23}
\newcommand{\FluxRatio}{43{,}100}
\newcommand{\NFifty}{4{,}103}
\newcommand{\KFifty}{0}
\newcommand{\MuFifty}{3.0}

\newcommand{\FFiftyHalf}{1.5\times10^{-3}}
\newcommand{\BLfCommon}{0.60}
\newcommand{\NGBTS}{1{,}005}
\newcommand{\KappaInj}{0.861}
\newcommand{\KappaCat}{0.899}
\newcommand{\KappaCatErr}{0.010}
\newcommand{\KappaCatN}{26}
\newcommand{\KappaSys}{5}
\newcommand{\EtaRange}{0.86$--$0.87}
\newcommand{\CtrlStdMax}{1.06}
\newcommand{\FPRawMin}{6.8\times10^{-5}}
\newcommand{\FPRawMax}{9.4\times10^{-5}}
\newcommand{\InjFour}{0.15}
\newcommand{\InjFive}{0.51}
\newcommand{\InjSix}{0.81}
\newcommand{\NInject}{24{,}000}
\newcommand{\NEpochTags}{4}
\newcommand{\NStarEpochs}{720{,}108}
\newcommand{\NCutoutsAll}{722{,}273}
\newcommand{\NBadCut}{2{,}165}
\newcommand{\NCutEpList}{208{,}573, 208{,}573, 207{,}717, and 97{,}410}
\newcommand{\NNoEpoch}{1{,}959}
\newcommand{\NNoEpochPct}{0.9}
\newcommand{\NNoEpochSouth}{335}
\newcommand{\NAllVeto}{18}
\newcommand{\NWithForced}{208{,}232}
\newcommand{\NCtrlAssoc}{16}
\newcommand{\NCoaddVeto}{5}
\newcommand{\CtrlAngleMin}{0}
\newcommand{\CtrlAngleMax}{3}
\newcommand{\PRankSteady}{0.67}
\newcommand{\BSteadyErrBoot}{0.4}
\newcommand{\BSnapErrBoot}{0.6}
\newcommand{\KHost}{3}
\newcommand{\BHost}{1.9}
\newcommand{\MuHost}{6.1}
\newcommand{\MuSteadyBz}{4.7}
\newcommand{\CatStarIn}{4}
\newcommand{\CatCtrlIn}{3.5}

\newcommand{\NCoaddStars}{208{,}214}
\newcommand{\NInactPhot}{32{,}667}
\newcommand{\FBzSeven}{5.3\times10^{-2}}
\newcommand{\FBzEight}{1.3\times10^{-3}}
\newcommand{\FBzNine}{1.5\times10^{-4}}
\newcommand{\FBzFloor}{2.3\times10^{-5}}
\newcommand{\FDeltaFloorBz}{1.1\times10^{-5}}

\newcommand{\FDutyOneBz}{3.8\times10^{-5}}
\newcommand{\FDutyThreeBz}{5.3\times10^{-5}}
\newcommand{\FDutyTenthBz}{1.2\times10^{-4}}
\newcommand{\AddDiffPct}{0.01}
\newcommand{\ReachIndBz}{1.4\times10^{6}}
\newcommand{\ReachIndEIRP}{2.8\times10^{15}}
\newcommand{\FloorL}{1.1\times10^{10}}
\newcommand{\FloorEIRP}{2.2\times10^{19}}
\newcommand{\FDutyOne}{2.9\times10^{-5}}
\newcommand{\FDutyThree}{4.2\times10^{-5}}
\newcommand{\FDutyTenth}{9.7\times10^{-5}}

\newcommand{\XcOwnAngleMin}{0}
\newcommand{\XcOwnAngleMax}{5}
\newcommand{\XcOwnPLow}{0.4232}
\newcommand{\XcOwnBErrBoot}{0.6}
\newcommand{\PctActTen}{73}

\newcommand{\PctActHundred}{9}
\newcommand{\NGCNS}{229{,}425}
\newcommand{\PctGCNS}{91}
\newcommand{\NGCNSmsw}{17{,}348}
\newcommand{\PctGCNSmsw}{7.6}
\newcommand{\NGCNSplx}{670}
\newcommand{\NGCNSfoot}{1{,}858}
\newcommand{\NContam}{642}
\newcommand{\PctContam}{0.3}

\newcommand{\NBrightDet}{3}
\newcommand{\NBrightUndet}{32}
\newcommand{\AlgolS}{32}
\newcommand{\ThetAurSNR}{6.1}
\newcommand{\XiTauSNR}{6.2}
\newcommand{\BrightLSirius}{5.4\times10^{5}}
\newcommand{\BrightLAltair}{5.1\times10^{5}}
\newcommand{\BrightLVega}{3.1\times10^{6}}
\newcommand{\BrightLFomalhaut}{3.5\times10^{6}}
\newcommand{\BrightUndetMaxSNR}{3.3}
\newcommand{\NInjAstro}{300}
\newcommand{\GoteChiStar}{0.9}
\newcommand{\GoteDofStar}{6}
\newcommand{\GoteChiStat}{9.5}
\newcommand{\GoteDofStat}{4}
\newcommand{\GotePStat}{0.051}
\newcommand{\GoteMotion}{1.34}
\newcommand{\LspmQLChiStat}{7.3}
\newcommand{\LspmQLChiCo}{18.3}
\newcommand{\LspmQLDof}{4}

\newcommand{\TycMotion}{0.56}
\newcommand{\TycPStat}{0.05}
\newcommand{\TycPCo}{0.06}

\newcommand{\CosN}{6{,}203}
\newcommand{\CosNmeas}{18{,}608}

\newcommand{\CosNcat}{10{,}830}
\newcommand{\CosRatioFaint}{1.10}
\newcommand{\CosRatioFaintErr}{0.04}
\newcommand{\CosLoBin}{20--50}
\newcommand{\CosLoRatio}{1.10}
\newcommand{\CosLoErr}{0.07}
\newcommand{\CosRatioMin}{1.03}
\newcommand{\CosRatioMax}{1.12}
\newcommand{\NPairsBeam}{1{,}867}
\newcommand{\NStarsPairs}{3{,}734}
\newcommand{\NCatOnlySingle}{8}
\newcommand{\NCatOnlySingleAct}{4}
\newcommand{\NCatOnlySingleElig}{0}
\newcommand{\KCoFour}{9}
\newcommand{\BCoFour}{4.6}
\newcommand{\KCoSeven}{1}
\newcommand{\BCoSeven}{0.5}
\newcommand{\KCoNegThree}{48}
\newcommand{\BCoNegThree}{51.9}

\newcommand{\NBrightMissing}{35}

\newcommand{\PChanceRxs}{1\times10^{-3}}
\newcommand{\PChanceEra}{4\times10^{-3}}
\newcommand{\NActive}{25{,}894}
\newcommand{\PctActive}{12.3}
\newcommand{\BSteadyErr}{0.4}
\newcommand{\BSnapErr}{0.5}
\newcommand{\NCtrlSteady}{10}
\newcommand{\NCtrlSnap}{18}
\newcommand{\MuSteadyMarg}{4.1}
\newcommand{\MuSnapMarg}{5.9}
\newcommand{\BSteadyAll}{11.3}
\newcommand{\BSnapAll}{15.4}
\newcommand{\BSteadyNoEmit}{1.3}
\newcommand{\BSnapNoEmit}{2.3}

\newcommand{\PSnapHigh}{0.39}
\newcommand{\KSteadyLxA}{2}
\newcommand{\BSteadyLxA}{1.3}
\newcommand{\KSnapLxA}{4}
\newcommand{\BSnapLxA}{2.3}

\newcommand{\NEligLxA}{33{,}398}
\newcommand{\KSteadyLxB}{1}
\newcommand{\BSteadyLxB}{1.3}
\newcommand{\KSnapLxB}{3}
\newcommand{\BSnapLxB}{2.1}

\newcommand{\NEligLxB}{31{,}612}
\newcommand{\KSteadyWithMult}{7}
\newcommand{\BSteadyWithMult}{1.6}
\newcommand{\KSnapWithMult}{10}
\newcommand{\BSnapWithMult}{2.8}
\newcommand{\PSteadyWithMult}{1\times10^{-3}}
\newcommand{\PSnapWithMult}{6\times10^{-4}}
\newcommand{\NEligWithMult}{46{,}650}

\newcommand{\NElig}{32{,}985}
\newcommand{\PctElig}{16}

\newcommand{\NExclHot}{1{,}818}
\newcommand{\NAFGKMult}{13{,}665}
\newcommand{\PctAFGKMult}{29}
\newcommand{\NEnv}{174{,}955}
\newcommand{\NEnvM}{161{,}290}

\newcommand{\NExclAll}{2{,}251}

\newcommand{\NEBall}{133}

\newcommand{\XLimRxsHi}{-2.9}
\newcommand{\XLimRxsPctFour}{51}

\newcommand{\XLimEraPctFour}{83}
\newcommand{\PctEligAFGKTen}{64}
\newcommand{\PctMultAFGKTen}{33}
\newcommand{\PctEligAFGKHundred}{67}
\newcommand{\PctMultAFGKHundred}{30}

\newcommand{\LDGCVn}{4.5\times10^{9}}
\newcommand{\NAFGKTen}{44}
\newcommand{\NOldEligAFGKTen}{5}
\newcommand{\PctOldEligAFGKTen}{11}

\newcommand{\PctOldEligAFGKHundred}{82}
\newcommand{\MaxMdwL}{4.5\times10^{9}}
\newcommand{\MaxStarL}{5.9\times10^{9}}
\newcommand{\MaxMdwSNR}{6.3}
\newcommand{\MaxMdwSigRatio}{10}
\newcommand{\NextMdwL}{2.3\times10^{9}}
\newcommand{\MaxStarName}{AE Aqr}
\newcommand{\EcfHR}{-0.44}
\newcommand{\NEligRxsOnly}{1{,}632}
\newcommand{\NEcfZero}{113}
\newcommand{\NEcfHard}{175}
\newcommand{\RxsDepthCR}{0.033}
\newcommand{\RxsMedCR}{0.029}

\newcommand{\PctXUncHundred}{44}
\newcommand{\PctRuweBright}{32}
\newcommand{\PctRuweFaint}{25}
\newcommand{\EpsEriRUWE}{2.7}

\newcommand{\MaxForcedML}{4.5\times10^{9}}
\newcommand{\NAFGKCand}{11}
\newcommand{\NAFGKCandMult}{8}

\newcommand{\NCandAFGKExcl}{60}
\newcommand{\NCandM}{90}

\newcommand{\NEnvPhot}{173{,}319}
\newcommand{\FBzElig}{1.5\times10^{-4}}

\newcommand{\FBzAtEnv}{4.0\times10^{-5}}
\newcommand{\FAllAtEnv}{3.4\times10^{-5}}
\newcommand{\NEnvSteady}{173{,}315}
\newcommand{\ReachMall}{2.9\times10^{6}}
\newcommand{\TycXLim}{-3.6}
\newcommand{\TycDist}{93.5}
\newcommand{\KMall}{72}
\newcommand{\KMallSec}{16}
\newcommand{\NMallCoadd}{159{,}937}
\newcommand{\MuMall}{87.6}
\newcommand{\FMallSeven}{1.6\times10^{-1}}
\newcommand{\NeffMallSeven}{563}
\newcommand{\FMallEight}{4.9\times10^{-3}}
\newcommand{\NeffMallEight}{18{,}061}
\newcommand{\FMallNine}{5.5\times10^{-4}}
\newcommand{\NeffMallNine}{159{,}550}
\newcommand{\MuWithMult}{13.1}
\newcommand{\NWithMultSteady}{46{,}206}
\newcommand{\FBzWithMult}{2.8\times10^{-4}}
\newcommand{\NSecEnv}{17}
\newcommand{\NSecEnvEp}{59}
\newcommand{\LSecEnvMax}{1.0\times10^{9}}
\newcommand{\KXc}{0}
\newcommand{\NXc}{20{,}421}
\newcommand{\NeffXcNine}{20{,}371}
\newcommand{\FXcNine}{1.47\times10^{-4}}
\newcommand{\FBzNineTwo}{1.46\times10^{-4}}
\newcommand{\FAllElig}{1.2\times10^{-4}}
\newcommand{\FBzEnvKZero}{1.5\times10^{-5}}
\newcommand{\NCandActEv}{149}

\newcommand{\PSteadyHigh}{0.71}
\newcommand{\TycSNR}{7.8}
\newcommand{\TycFieldSimR}{1}

\newcommand{\TycFieldSimNear}{26}
\newcommand{\TycFieldGaiaNear}{6.5}
\newcommand{\TycNvssName}{NVSS J202939$+$460047}
\newcommand{\TycNvssSep}{17}
\newcommand{\TycNvssS}{2.5}
\newcommand{\TycNvssES}{0.5}
\newcommand{\TycNvssSig}{2.6}
\newcommand{\TycProjAU}{170}
\newcommand{\TycFieldB}{4}
\newcommand{\TycG}{11}
\newcommand{\NEligSteady}{32{,}666}
\newcommand{\NCandXcat}{137}
\newcommand{\NCandSimbadOnly}{12}
\newcommand{\NCandTX}{96}
\newcommand{\NCandTFl}{30}
\newcommand{\NCandTEm}{15}
\newcommand{\NCandTBY}{9}
\newcommand{\NCandTRS}{8}
\newcommand{\NCandTY}{4}
\newcommand{\NCandTCP}{1}
\newcommand{\NCandStars}{161}

\newcommand{\KUnexpl}{3}

\newcommand{\KSteady}{1}
\newcommand{\BSteady}{1.3}
\newcommand{\MuSteady}{4.0}
\newcommand{\KSnap}{3}
\newcommand{\BSnap}{2.3}
\newcommand{\MuSnap}{5.8}
\newcommand{\NCandCoadd}{114}
\newcommand{\NCandEpoch}{139}

\newcommand{\NVetoStarEp}{76}
\newcommand{\NStarsVeto}{26}

\newcommand{\MedEpochs}{3}
\newcommand{\NEmitters}{211}
\newcommand{\NEmitPct}{0.1}
\newcommand{\SlimAllMed}{0.54}
\newcommand{\SlimAllLo}{0.45}
\newcommand{\SlimAllHi}{0.69}
\newcommand{\LlimAllMed}{4.0\times10^{8}}
\newcommand{\EIRPlimAllMed}{8.0\times10^{17}}

\newcommand{\SlimAnyMed}{1.12}
\newcommand{\SlimBestMed}{0.91}
\newcommand{\BarnardLAll}{2.3\times10^{5}}
\newcommand{\BarnardEIRPAll}{4.5\times10^{14}}

\newcommand{\BarnardNep}{3}
\newcommand{\EarthLeakAUAll}{16{,}300}

\newcommand{\NAstars}{768}
\newcommand{\FFiftyRatio}{4}
\newcommand{\FAllSeven}{4.5\times10^{-2}}

\newcommand{\FAllEight}{1.1\times10^{-3}}

\newcommand{\FAllNine}{1.2\times10^{-4}}

\newcommand{\FAllFloor}{2.0\times10^{-5}}
\newcommand{\NAllFloor}{205{,}981}
\newcommand{\FDeltaFloor}{8.5\times10^{-6}}

\newcommand{\ZuckSband}{15}
\newcommand{\ZuckL}{7.2\times10^{13}}
\newcommand{\ZuckEIRP}{1.4\times10^{23}}
\newcommand{\ZuckChan}{1.5\times10^{4}}
\newcommand{\ZuckMarginMed}{10^{5}}
\newcommand{\ZuckMarginWorst}{2\times10^{3}}
\newcommand{\LlimAllMax}{2.5\times10^{10}}
\newcommand{\BPRPlo}{0.64}
\newcommand{\BPRPhi}{2.23}
\newcommand{\NSolarSteady}{1}
\newcommand{\NSolarSingle}{45{,}360}

\newcommand{\NSolarActive}{991}
\newcommand{\NSolarInactive}{44{,}369}
\newcommand{\NSolarUnexpl}{2}
\newcommand{\FSolarSingle}{1.1\times10^{-4}}

\newcommand{\AllMSLAct}{1.9\times10^{6}}
\newcommand{\AllMSLErrAct}{1.7\times10^{5}}
\newcommand{\AllMSNRAct}{11.3}
\newcommand{\AllMNAct}{2{,}216}
\newcommand{\AllMSLIn}{4.3\times10^{5}}
\newcommand{\AllMSLErrIn}{1.1\times10^{5}}
\newcommand{\AllMSNRIn}{4.1}
\newcommand{\AllMNIn}{7{,}645}

\newcommand{\ActFracMForty}{22}
\newcommand{\InMmaxSNR}{4.2}
\newcommand{\ActAFGKmaxSNR}{6.3}
\newcommand{\InAFGKmaxSNR}{1.9}
\newcommand{\AllMSL}{8.1\times10^{5}}
\newcommand{\AllMSLErr}{9.2\times10^{4}}
\newcommand{\AllMSNR}{8.8}
\newcommand{\AllMN}{9{,}861}
\newcommand{\AllMNeffW}{3{,}650}
\newcommand{\AllMDmed}{0.084}
\newcommand{\AllMDmedErr}{0.014}
\newcommand{\AllMDmedSNR}{5.8}
\newcommand{\AllAFGKmaxSNR}{4.0}
\newcommand{\AllAFGKmaxShell}{60--80}
\newcommand{\NMShellsFour}{4}

\newcommand{\CandAstroNdet}{4}
\newcommand{\CandAstroOff}{1.8}
\newcommand{\CandAstroChi}{194}
\newcommand{\CandAstroDof}{8}
\newcommand{\CandAstroS}{1.44}
\newcommand{\TopTenMax}{4.6}

\newcommand{\ReachAllAFGK}{1.4\times10^{5}}
\newcommand{\FMeanSixAFGK}{0.30}
\newcommand{\FBratAllAFGK}{0.48}
\newcommand{\NeffSixAFGK}{77}
\newcommand{\WtenSixAFGK}{37}
\newcommand{\NeffBratAFGK}{39}
\newcommand{\WtenBratAFGK}{58}

\newcommand{\FMedSevenAFGK}{0.084}
\newcommand{\FMedNineAFGK}{0.016}
\newcommand{\StackCrossAFGK}{8.3\times10^{6}}

\newcommand{\NCandSubCoadd}{47}
\newcommand{\GoteSnrLo}{4.5}
\newcommand{\GoteSnrHi}{6.9}

\newcommand{\CovAFGKmin}{0.88}
\newcommand{\NCovTrials}{200}
\newcommand{\CovNangles}{8}
\newcommand{\ReachRatioAFGK}{10}

\newcommand{\SnapOnlyLo}{5.0}
\newcommand{\SnapOnlyHi}{5.5}

\newcommand{\GOTEL}{4.6\times10^{7}}

\newcommand{\GOTEd}{24}
\newcommand{\GOTEdMG}{0.7}

\newcommand{\GOTEsep}{0.2}
\newcommand{\CandAstroScoadd}{0.53}
\newcommand{\NMSecNat}{18}
\newcommand{\MSecLlo}{5.3\times10^{6}}
\newcommand{\MSecLhi}{1.5\times10^{9}}
\newcommand{\XcSEcovTwo}{44}
\newcommand{\XcSEcovThree}{46}
\newcommand{\XcKappaRange}{0.83--0.87}
\newcommand{\XcSEratio}{1.13}
\newcommand{\XcSEratioErr}{0.01}
\newcommand{\XcSEratioN}{28}
\newcommand{\XcSEratioRange}{1.07--1.16}

\newcommand{\XcSEoverQL}{0.975}
\newcommand{\XcSEoverQLErr}{0.006}
\newcommand{\XcSEoverQLN}{17}

\newcommand{\XcAstSE}{0\farcs05--0\farcs19}
\newcommand{\XcAstQL}{0\farcs10--0\farcs21}
\newcommand{\XcConcQL}{85}
\newcommand{\XcConcQLN}{127}
\newcommand{\XcConcQLpct}{67}
\newcommand{\XcConcSE}{69}
\newcommand{\XcConcSEN}{77}
\newcommand{\XcConcSEpct}{90}
\newcommand{\XcCatLtFive}{50}
\newcommand{\XcCatLtFiveOff}{37}
\newcommand{\XcCatChance}{52}

\newcommand{\XcCandCat}{70}
\newcommand{\XcOwnKs}{2}
\newcommand{\XcOwnBs}{3.0}
\newcommand{\XcOwnMus}{4.4}
\newcommand{\XcOwnFBz}{1.9\times10^{-4}}
\newcommand{\XcOwnMuBz}{6.3}
\newcommand{\XcOwnFBzPct}{33}
\newcommand{\XcOwnKe}{4}
\newcommand{\XcOwnBe}{4.9}
\newcommand{\XcOwnMue}{5.5}
\newcommand{\XcUnionKs}{1}
\newcommand{\XcUnionBs}{1.3}

\newcommand{\XcUnionFBz}{1.5\times10^{-4}}
\newcommand{\XcUnionMuBz}{4.7}

\newcommand{\XcUnionKe}{3}
\newcommand{\XcUnionBe}{2.1}

\newcommand{\XcUnionPe}{0.36}
\newcommand{\XcOwnAdded}{8}
\newcommand{\XcOwnAddedUnexpl}{1}

\newcommand{\XcMuSnapChange}{+1}
\newcommand{\XcMuSteadyChange}{{<}1}
\newcommand{\XcCatOnlyN}{12}
\newcommand{\XcCatOnlyMulti}{4}
\newcommand{\XcMedN}{152{,}511}
\newcommand{\XcMedDet}{46}
\newcommand{\XcMedSecure}{20}
\newcommand{\XcMedSecureTested}{28}
\newcommand{\XcMedExplained}{11}

\newcommand{\XcMedUnexplained}{0}

\newcommand{\XcMedCatonly}{6}

\newcommand{\XcMedNone}{9}

\newcommand{\XcMedChance}{22}
\newcommand{\XcMedRms}{94}
\newcommand{\XcNCat}{8}
\newcommand{\XcSEcons}{13}
\newcommand{\VCepSep}{0.65}
\newcommand{\VCepRel}{0.01}
\newcommand{\VCepLaterSep}{0.14}
\newcommand{\XcMoveAName}{BD+44 2051B}
\newcommand{\XcMoveASpan}{23}
\newcommand{\XcMoveACo}{32}
\newcommand{\XcMoveASt}{14452}
\newcommand{\XcMoveADof}{2}
\newcommand{\XcMoveBName}{G 272-61A}
\newcommand{\XcMoveBSpan}{8}
\newcommand{\XcMoveBCo}{2}
\newcommand{\XcMoveBSt}{964}

\newcommand{\XcMedNoneSepMin}{1.9}
\newcommand{\XcMedNoneSepMax}{2.3}

\shorttitle{A wideband VLASS technosignature search}
\shortauthors{Croft}
\submitjournal{AJ}

\makeatletter
\newcommand{\asec}[1]{\expandafter\asec@i#1..\@nil}
\def\asec@i#1.#2.#3\@nil{\ifx\relax#2\relax#1\arcsec\else#1\farcs#2\fi}
\makeatother
\newcommand{\Lnu}{L_\nu}
\newcommand{\WHz}{\mathrm{W\,Hz^{-1}}}

\newcommand{\BLBerkeley}{Breakthrough Listen, University of California, Berkeley, 501 Campbell Hall \#3411, Berkeley, CA 94720, USA}
\newcommand{\seti}{SETI Institute, 339 Bernardo Ave, Suite 200, Mountain View, CA 94043, USA}
\newcommand{\BLOxford}{Breakthrough Listen, University of Oxford, Denys Wilkinson Building, Keble Road, Oxford OX1 3RH, UK}

\begin{document}

\title{A Wideband Technosignature Search of \NSample\ Nearby Main-sequence Stars with the VLA Sky Survey}

\author[orcid=0000-0003-4823-129X, sname='Croft']{Steve Croft} 
\affiliation{\BLOxford}
\affiliation{\BLBerkeley}
\affiliation{\seti}
\email[show]{scroft@berkeley.edu}
\correspondingauthor{Steve Croft}

\begin{abstract}
Searches for radio technosignatures have concentrated on narrowband signals. We present a complementary search for excess broadband 2--4\,GHz emission from \NSample\ Gaia DR3 main-sequence stars within 100\,pc in all epochs of the VLA Sky Survey (VLASS; 2017--2026). Cross-matching with the Epoch-1 catalog yields \NSecure\ high-reliability positional associations (versus \NFalseSecure\ expected by chance). All but one are known natural emitters that, where measurable, move with their stars; the remaining source, near LSPM~J1943+5337, is not at the star. Because stellar flares can mimic transmitters in continuum flux, a detection counts as a possible transmitter at any luminosity only at the \NElig\ apparently single AFGK stars outside known radio-emitter classes, and at M dwarfs and AFGK multiples only above $10^{10}\,\WHz$. Forced photometry of \NWithForced\ stars in all epochs finds \KSteady\ steady-test detection and \KHost\ stars detected in at least one epoch among the eligible stars, against \BSteady\ and \BHost\ expected by chance. The median per-star limit on steady emission is $\Lnu<\LlimAllMed\,\WHz$. Without background subtraction, fewer than $\FBzNine$ of eligible AFGK stars host a steady transmitter above $10^9\,\WHz$, and fewer than $\FBzFloor$ of all stars outside the known-emitter classes host one above $\sim\FloorL\,\WHz$; counting every M-dwarf detection as a possible transmitter, fewer than $\FMallNine$ of M dwarfs host one above $10^9\,\WHz$. Stacks detect population-averaged emission from M dwarfs, concentrated in those with activity evidence. Continuum surveys thus constrain broadband, modulated, or rapidly drifting signals that narrowband pipelines do not search for.
\end{abstract}


\section{Introduction} \label{sec:intro}

Radio searches for technosignatures have, since their inception, concentrated on narrowband emission \citep[e.g.,][]{Tarter2001}. A signal much narrower than any known natural maser line is an unambiguous marker of technology, and narrowband searches with modern spectrometers reach impressive sensitivity: the Breakthrough Listen (BL) survey of nearby stars \citep{Isaacson2017} placed limits of order $10^{13}$\,W \citep[including the correction of][]{Choza2024} on the equivalent isotropic radiated power (EIRP) of drifting narrowband transmitters around 1327 targets drawn from a nominal 50\,pc sample \citep{Price2020}, and archival and commensal programs have since extended such searches to many more targets \citep[e.g.,][]{Tremblay2025}. \citet{Painter2025} took a step beyond the drifting-tone paradigm, searching thousands of stars in the BL Green Bank Telescope archive for departures from Gaussian kurtosis within 2.86\,kHz spectral blocks rather than matching a template of a single unresolved, Doppler-drifting line---a stepping stone toward searches for broader and more complex signals---and \citet{Gajjar2022} searched 1883 stars for artificially dispersed broadband pulses.

Terrestrial transmitters are increasingly broadband---spread-spectrum, frequency-hopped, chirped radar and multi-carrier modulations dominate modern usage---and a transmitter whose power is spread over kHz to GHz is effectively invisible to pipelines tuned to Hz-wide drifting lines. Earth's own leakage has shifted from a modest number of powerful broadcast carriers, which would be detectable from nearby stars \citep{Sullivan1978,LoebZaldarriaga2007,Sheikh2025}, to huge numbers of lower-power transmitters; \citet{Saide2023} simulated the leakage from the global network of mobile-communication towers as it would appear from nearby stellar systems. \citet{Garrett2026} argued that the aggregate output of such distributed infrastructure would appear to a distant observer as faint, compact continuum emission---a broadband radio technosignature (BRaT)---and, scaling Earth's present aggregate leakage ($\Lnu\sim400\,\WHz$ at L band) linearly with power consumption, estimated $\Lnu\sim4\times10^5\,\WHz$ for a Kardashev Type~I civilization \citep{Kardashev1964}. \citeauthor{Garrett2026} proposed a hierarchical search in which wide-field continuum surveys identify compact sources coincident with nearby stars, and very long baseline interferometry (VLBI) then tests them for high brightness temperature, negligible polarization, non-uniform spectra, scintillation, and co-motion with the host star, and noted that existing surveys can already place the first constraints on BRaT prevalence. \citet{Zuckerman2026} reached a related conclusion from the transmitter's side: a nearby civilization seeking contact would beam continuous, high-power signals toward a few hundred Earth-like planets, so that power starvation, and with it the need for Hz-wide channels, no longer applies. He noted that general-purpose broadband surveys such as VLASS are sensitive enough to detect such a signal from old solar-type stars within 200\,pc, but considered it premature to derive a quantitative limit from them.

Nevertheless, existing wide-area radio continuum surveys can provide interesting constraints, as they already have for Kardashev Type~III civilizations in other galaxies \citep{Garrett2015}. Such surveys integrate the total power in a broad band irrespective of modulation, drift rate, or bandwidth, and they cover a large fraction of the sky to a uniform depth, so that every star in the survey area is observed. The cost is that a stellar continuum source is not intrinsically artificial: active stars emit gyrosynchrotron and coherent radio emission \citep{GudelBenz1993,Callingham2021,Vedantham2022}, and the overwhelming majority of mJy radio sources are extragalactic and can align with a foreground star by chance. A continuum technosignature search is therefore a search for \emph{unexplained} radio emission, and its power lies in the care with which natural explanations and chance alignments are eliminated \citep{Garrett2026}.

Wide-field surveys have recently turned radio stars from individual curiosities into a population. \citet{Driessen2023} used Gaia \citep{GaiaMission} proper motions to identify stars whose radio counterparts in the Faint Images of the Radio Sky at Twenty-cm survey \citep[FIRST;][]{Becker1995}, the VLA Sky Survey \citep[VLASS;][]{Lacy2020}, and the Rapid ASKAP Continuum Survey \citep[RACS;][]{McConnell2020} moved with them, and the Sydney Radio Star Catalogue now compiles more than 800 radio-detected stars \citep{Driessen2024}. \citet{Yiu2024} used detections in VLASS and the LOFAR Two-metre Sky Survey \citep[LoTSS;][]{Shimwell2022} to show that radio emission traces stellar magnetic activity, and \citet{De2024} cross-matched VLASS with Gaia DR3 to characterize the radio-star population within 500\,pc. \citet{Davis2026} detected VLASS counterparts to six of 150 superflaring solar-type stars, and multi-epoch ASKAP sampling shows that at least $\sim$10\% of M dwarfs produce radio bursts above $10^{15}$\,erg\,s$^{-1}$\,Hz$^{-1}$, with a burst duty cycle below 8.5\% \citep{Pritchard2024}. Each survey approach---selection by variability, circular polarization, or proper motion---carries its own bias \citep{Matthews2025}; forced photometry at the position of every star in a sample selected independently of radio properties avoids these selections. These studies set out to find and understand radio stars. Our aim is complementary: for a nearly complete sample of the main-sequence stars within 100\,pc (Section~\ref{sec:gaia}), we quantify the probability that each radio association is genuine, require a natural explanation for every genuine association, and convert the non-detections, individually and in stacks, into limits on artificial emission.

Here we carry out our search using all VLASS epochs observed to date (2017--2026). VLASS covers the sky north of $\delta=-40^\circ$ at 2--4\,GHz with $\sim$2\farcs5 resolution and $\sim$0.12--0.15\,mJy\,beam$^{-1}$ noise per epoch; we combine it with Gaia DR3 \citep{GaiaDR3}. We cross-match all main-sequence stars within 100\,pc against the Epoch-1 VLASS Quick Look component catalog \citep{Gordon2020,Gordon2021}, quantify chance alignments with offset catalogs, classify every association, and follow up the one unexplained source with archival optical imaging and multi-epoch VLASS astrometry. We then measure every star in every epoch by forced photometry, define by stellar type and luminosity where a detection could be recognized as a possible transmitter, calibrate the chance background at control positions, and check the results against independent per-epoch catalogs. Finally, we derive limits on individual stars and on the population, and stack the images of all individually undetected stars. Section~\ref{sec:data} describes the data, Section~\ref{sec:methods} the methods, Section~\ref{sec:results} the results, and Section~\ref{sec:discussion} compares them with narrowband searches and with broadband leakage benchmarks. All code and data needed to reproduce this work are archived and are publicly available (Section~\ref{sec:data_avail}).

\section{Data} \label{sec:data}

\subsection{Nearby main-sequence stars from Gaia DR3} \label{sec:gaia}

We selected the \NGaia\ Gaia DR3 sources with parallax $\varpi>10$\,mas, $\varpi/\sigma_\varpi>10$, and $\delta>-45^\circ$. Of these, \NFinite\ have a $G_{BP}-G_{RP}$ color. We defined the main sequence empirically as the iteratively clipped median absolute magnitude $M_G$ in 0.1\,mag bins of $G_{BP}-G_{RP}$, after coarse cuts removing white dwarfs and giants, and retained stars within $-1.5<\Delta M_G<+1.0$\,mag of this ridge. The asymmetric window keeps unresolved binaries (up to 0.75\,mag over-luminous for equal-mass pairs) and excludes the white-dwarf sequence and the spurious faint sources below the main sequence that are known to contaminate Gaia nearby-star samples \citep{Smart2021}. This leaves \NMS\ stars, of which \NMSdec\ lie at $\delta>-41^\circ$ and \NSample\ fall within the VLASS Epoch~1 footprint. We divide the sample at $G_{BP}-G_{RP}=1.84$, close to the K7/M0 boundary \citep{Pecaut2013}, into AFGK stars (including \NAstars\ A stars bluer than $G_{BP}-G_{RP}=0.33$) and M dwarfs. Compared with the Gaia Catalogue of Nearby Stars \citep[GCNS;][]{Smart2021}, which assigns every EDR3 source a probability of lying within 100\,pc, the sample contains \PctGCNS\% of the \NGCNS\ objects that GCNS places within 100\,pc at $\delta>-41^\circ$ and does not classify as white dwarfs. Most of the others (\NGCNSmsw, \PctGCNSmsw\%) fall outside our main-sequence window, for example strongly over-luminous unresolved binaries and stars with problematic photometry, \NGCNSfoot\ lie outside the VLASS footprint, and only \NGCNSplx\ fail our parallax cuts; conversely, \NContam\ sample stars (\PctContam\%) are not placed within 100\,pc by GCNS. The brightest stars are discussed in Section~\ref{sec:caveats}. \NPairsBeam\ pairs of sample stars (\NStarsPairs\ stars, mostly wide binaries) are closer than 2\farcs5, so their measurements are not independent; they make up less than 2\% of the sample.

\subsection{VLASS catalogs and images} \label{sec:vlass}

VLASS Epoch~1 was observed in two halves (1.1 and 1.2) between \EpochMin\ and \EpochMax\ \citep{Lacy2020,Kimball2026}. We use the CIRADA Quick Look (QL) component catalog \citep{Gordon2020,Gordon2021}, which contains \NComp\ components extracted with PyBDSF \citep{Mohan2015} from \NSubtiles\ $1\,\mathrm{deg}^2$ QL images. We use all \NUsedComp\ non-duplicate components with a fitted Gaussian (not only the \NMainSample\ in the recommended ``main sample''), so that faint and low-quality components near stars are also considered. QL peak intensities are low by $\sim$15\% in half-epoch 1.1, whose observations were affected by antenna pointing errors, and by $\sim$8\% in later half-epochs because of residual phase errors, while integrated flux densities are low by $\sim$10\% in 1.1 and accurate to $\sim$3\% otherwise \citep{Kimball2026}. We therefore divide peak intensities and image noise by $b_{\rm pk}=0.85$ (half-epoch 1.1) or 0.92 (later half-epochs), and catalog total flux densities by 0.90 (1.1) or 1.

The Epoch-1 QL images were made without correcting for the w-terms, which displaces positions by up to $\sim$1\arcsec\ at low elevation; the correction was later applied to the image metadata \citep{Kimball2026}, but the Epoch-1 catalog predates it. Matching \EOnePairs\ compact (deconvolved major axis $<1\farcs5$), isolated, S/N\,$>20$ Epoch-1 components to the Epoch-3 QL lists, whose images are corrected, we find Epoch-1 declinations too large by a median of \asec{\EOneDNsouth} at $\delta<-20^\circ$, falling to too small by \asec{\EOneDNnorth} at $\delta>60^\circ$, with right ascensions offset by $\lesssim0\farcs1$. We subtract from every Epoch-1 position the median offset of its image (for \EOneNsub\ images with at least 10 reference components), of its observing tile, which is observed in a single pass (\EOneNtile\ images), or otherwise the running median at its declination (\EOneNdec\ images), and carry the uncertainty of the adopted offset (median \asec{\EcorrMed} per axis) into the positional errors below.

The CIRADA catalog also lists, for each component, the distance to the nearest NVSS \citep{Condon1998} and FIRST \citep{Becker1995,Helfand2015} source, which we use to test whether radio sources move with their associated stars. For image-based analyses we retrieved QL images for all seven VLASS half-epochs observed to date (1.1--4.1, 2017--2026) and Single Epoch images for epochs 1.2 and 3.2 from the Canadian Astronomy Data Centre (CADC); the CADC copies of the QL images include the correction of image positions for the neglected w-terms \citep{Kimball2026}. Where the CADC metadata lack an observation date we took it from the image header. For an independent check of the later epochs (Section~\ref{sec:catcheck}) we also use component catalogs made with PyBDSF from full images: the NRAO interim QL component lists for half-epochs 3.1, 3.2, and 4.1 and for the median stack of the Epoch 1--3 QL images \citep{Lacy2025}, and the NRAO component lists from the Single Epoch (SE) continuum images of half-epochs 2.1--3.2 (versions of 2026 June). The SE images are made with self-calibration, masking, and $0\farcs6$ pixels and are of higher fidelity than QL \citep{Kimball2026}, but so far cover only part of the sky: \XcSEcovTwo\% and \XcSEcovThree\% of our Epoch-2 and Epoch-3 star-epochs.

\subsection{Ancillary data} \label{sec:ancillary}

For every star with a VLASS component within 2\farcs5 we retrieved the nearest ROSAT 2RXS \citep[within 40\arcsec;][]{Boller2016} and eROSITA eRASS1 \citep[within 20\arcsec;][]{Merloni2024} X-ray sources via VizieR \citep{Ochsenbein2000}, and SIMBAD \citep{Wenger2000} identifications and object types, matched by Gaia DR3 identifier rather than by position; activity evidence and technosignature eligibility for every sample star, which use the full X-ray catalogs and the Gaia DR3 eclipsing-binary classifications \citep{Mowlavi2023}, are described in Section~\ref{sec:candidates}. To test the flux scale below the detection threshold we use the VLA-COSMOS 3\,GHz catalog \citep{Smolcic2017}, retrieved from HEASARC. For the one unexplained secure association we also used Digitized Sky Survey (POSS-I and POSS-II) plates and Pan-STARRS1 images and catalogs \citep{Chambers2016,Flewelling2020}.

\section{Methods} \label{sec:methods}

\subsection{Epoch propagation and cross-matching} \label{sec:xmatch}

Many nearby stars move by several arcseconds between the Gaia reference epoch (J2016.0) and the VLASS observations. We assigned each star to the QL image whose center is nearest, and propagated its position to that image's observation date using its Gaia proper motion and parallax (the parallactic displacement reaches 0\farcs55 for Barnard's star); for the final reliability analysis each star--component pair was re-propagated to the date of the image in which the component was measured. We then found all components within 10\arcsec\ of each star.

\begin{figure}
\centering
\includegraphics[width=0.6\textwidth]{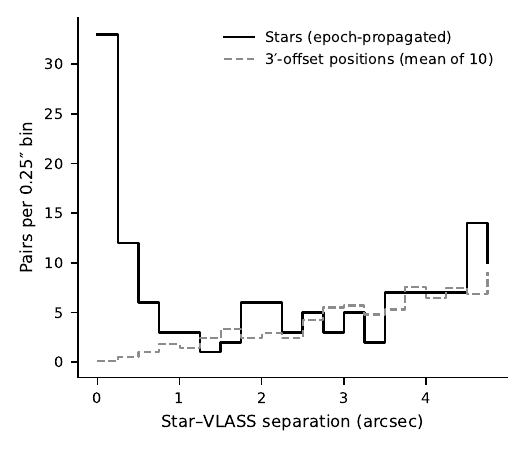}
\caption{Separations between epoch-propagated stars and VLASS components (solid) compared with the mean of ten 3\arcmin-offset realizations (dashed). The excess within 1\arcsec\ is due to genuine stellar associations; beyond $\sim$1\arcsec\ the two distributions agree.\label{fig:sep}}
\end{figure}

To measure the rate of chance alignments directly, we repeated the match ten times after displacing every star by 3\arcmin\ in a random direction. Figure~\ref{fig:sep} compares the separation distributions: there are \NRealOne\ star--component pairs within 1\arcsec, against \NOffOne\ per offset realization, while beyond $\sim$1\arcsec\ the two distributions agree.

\subsection{Positional reliability} \label{sec:reliability}

Following the likelihood-ratio approach to source identification \citep{SutherlandSaunders1992}, we modeled the separations $r_i$ of all pairs within $R_{\max}=10\arcsec$ as a mixture of genuine associations, with a two-dimensional Gaussian offset distribution, and background sources with a uniform surface density $\lambda$ per star fixed from the offset catalogs:
\begin{equation}
p_{\rm true}(r_i)=f\,\frac{r_i}{\sigma_i^2}\exp\!\left(-\frac{r_i^2}{2\sigma_i^2}\right),\qquad
p_{\rm bg}(r_i)=2\pi\lambda r_i,\qquad
\sigma_i^2=(k\,\sigma_{{\rm cat},i})^2+\sigma_{{\rm corr},i}^2+\sigma_{\rm sys}^2,
\end{equation}
where $\sigma_{{\rm cat},i}$ is the per-axis catalog positional error, scaled by a factor $k$ because PyBDSF errors are optimistic, $\sigma_{{\rm corr},i}$ the uncertainty of the astrometric correction, $\sigma_{\rm sys}$ a residual systematic floor, and $f$ the fraction of stars genuinely detected. Maximizing the extended likelihood gives $k=\ErrScaleK$, $\sigma_{\rm sys}=\asec{\SigSys}$ per axis, and $\sim$\NTrue\ genuine associations. The reliability of each pair is $R_i=p_{\rm true}/(p_{\rm true}+p_{\rm bg})$. We call pairs with $R>0.5$ secure associations, a statement about positional reliability only, not about the origin of the emission: there are \NSecure\ (\NRelNine\ with $R>0.9$), among which $\sum(1-R)=\NFalseSecure$ chance alignments are expected. One known radio star, the RS~CVn binary V376~Cep, falls below the threshold ($R=\VCepRel$): its corrected Epoch-1 position lies \asec{\VCepSep} from the star, whereas the Epoch-3 QL catalog places it \asec{\VCepLaterSep} away (Section~\ref{sec:catcheck}). Its Epoch-1 component is then treated as an unrelated source, so the star is confusion-limited and its limit is its measured flux density. The model assumes emission centered on the Gaia position: emission from a companion more than $\sim$1\arcsec\ away has a low $R$ and also lies outside the forced-photometry measurement at the star (Section~\ref{sec:limits_method}), so this search does not constrain emission from wide companions.

\subsection{Classification of associations} \label{sec:classify}

For each secure association ($R>0.5$) we recorded two kinds of evidence separately. Activity evidence is an X-ray counterpart (in 2RXS or eRASS1, or a SIMBAD X-ray-source type, since eRASS1 covers only half the sky) or a SIMBAD object type indicating flares, BY~Dra or RS~CVn variability, H$\alpha$ emission (emission-line star; on the main sequence this indicates chromospheric activity), youth (T~Tauri star or young stellar object), accretion (cataclysmic variable), or a magnetic chemically peculiar star. The same criteria are applied to the forced detections (Section~\ref{sec:candidates}). Multiplicity evidence is a SIMBAD binary object type, a Gaia non-single-star solution, RUWE\,$>1.4$ \citep{Lindegren2021}, or over-luminosity $\Delta M_G<-0.5$\,mag relative to the main-sequence ridge. Multiplicity is not itself a radio-emission mechanism; we record it because close binaries (RS~CVn and BY~Dra systems) are over-represented among radio-bright stars and because it flags systems in which the emission may arise in a companion.

Each star was then assigned one class, in order of precedence: cataclysmic variable; magnetic chemically peculiar star; RS~CVn binary; young star; BY~Dra variable; flare star; emission-line star (all from SIMBAD object types); X-ray-active star; otherwise unexplained. Multiplicity evidence alone does not explain an association, the same rule as for the forced detections (Section~\ref{sec:candidates}). Two stars typed in SIMBAD as RS~CVn or BY~Dra variables, PW~And and LQ~Hya, are young single dwarfs \citep{LopezSantiago2003,Lehtinen2012} and are classed as young stars. For stars with X-ray counterparts we computed $\log(L_X/\Lnu)$, using the eRASS1 0.2--2.3\,keV flux where available and otherwise the 2RXS count rate with a coronal energy conversion factor of $6\times10^{-12}$\,erg\,cm$^{-2}$\,count$^{-1}$ (0.1--2.4\,keV; the factor $(8.31+5.30\,{\rm HR1})\times10^{-12}$ of \citealt{Schmitt1995} for a soft corona with ${\rm HR1}=\EcfHR$), for comparison with the G\"udel--Benz relation, $L_X/\Lnu\approx10^{15.5}$\,Hz \citep{GudelBenz1993}.

Unexplained associations were then tested for being background sources. (1) A radio source that coincides with an NVSS or FIRST source at the VLASS position, although the star has since moved by more than twice that offset, did not move with the star and is therefore a background source. (2) For the one source surviving this test we compared archival optical images taken when the star was displaced from the radio position, and measured the radio position in three VLASS epochs relative to the star's Gaia track (Section~\ref{sec:lspm}).

\subsection{Forced photometry in all VLASS epochs} \label{sec:forced}

Beyond the Epoch-1 catalog, we measured every star in every VLASS epoch by forced photometry. For each of the \NEpochTags\ epochs (half-epochs 1.1 and 1.2, 2.1 and 2.2, 3.1 and 3.2, and 4.1) we extracted from CADC a $41\arcsec\times41\arcsec$ cutout of the QL image whose center is nearest to the star, centered on its proper-motion-propagated position: \NCutEpList\ cutouts in Epochs 1--4 (\NCutoutsAll\ in all; epoch 4.1 covers half the sky). A star has no cutout in an epoch if its 41\arcsec\ box would cross the edge of the nearest image (we did not fall back to an overlapping neighbor) or, for a few hundred stars, if no image of that epoch lies within 0\fdg72. Discarding \NBadCut\ cutouts with blanked pixels at the star leaves \NStarEpochs\ usable star-epochs (median \MedEpochs\ per star) for \NWithForced\ stars. The remaining \NNoEpoch\ sample stars (\NNoEpochPct\%; \NNoEpochSouth\ of them at the southern survey limit, $\delta<-38\arcdeg$) have no usable epoch; this loss depends only on position within the image grid, not on radio properties. A further \NAllVeto\ stars are vetoed for confusion (Section~\ref{sec:candidates}) in every epoch, leaving \NCoaddStars\ with a coadd; \NInactPhot\ of these are technosignature hosts eligible at any luminosity and \NEnvPhot\ are eligible only at high luminosity (Section~\ref{sec:candidates}); after the coadd-unit veto of the steady test, \NEligSteady\ and \NEnvSteady\ of them, \NAllFloor\ in all, remain in it. The flux density $S$ is the image interpolated bilinearly at the star's position including parallax, and the noise $\sigma$ is the robust (median absolute deviation) rms of the cutout beyond 10\arcsec\ from the star, scaled by a factor $\eta=\EtaRange$ so that $S/\sigma$ has unit scatter at blank positions: eight control positions 14\arcsec\ from each star, measured with the same estimator. $\eta$ is below unity because bilinear interpolation averages neighboring, partly correlated pixels, and because faint sources and sidelobes slightly inflate the cutout rms; we therefore measure it rather than model it. Bilinear interpolation of the $\sim2\farcs9\times2\farcs2$ beam sampled at 1\arcsec\ recovers a fraction $\kappa$ of the true peak. Injecting \NInject\ point sources into the Epoch-1 images at the positions of randomly chosen stars, each with the beam of the nearest Epoch-1 catalog component as a proxy for the local restoring beam (the same proxy is used for the injection tests of later epochs), and re-measuring them gives $\kappa=\KappaInj$, which we adopt; the catalog peaks of the brightest secure associations give $\kappa=\KappaCat\pm\KappaCatErr$ ($N=\KappaCatN$), and we treat the \KappaSys\% difference as a systematic uncertainty in the flux scale (the adopted value gives the more conservative limits). Flux densities are divided by the QL peak bias $b_{\rm pk}$ of their half-epoch (Section~\ref{sec:vlass}); as a check, the flux scales of compact reference sources relative to the Epoch-1 catalog (Section~\ref{sec:multiepoch}) agree with the ratios of these biases to within \FsAgree\%. The injected sources are recovered at $\ge5\sigma$ in \InjFour, \InjFive, and \InjSix\ of Epoch-1 cases at true $S/\sigma=4$, 5, and 6, as expected for Gaussian noise (Section~\ref{sec:stacks}). At the control positions $S/\sigma$ has zero mean and a scatter of at most \CtrlStdMax, but its tail above $5\sigma$, $\FPRawMin$--$\FPRawMax$ per trial before the confusion veto, is far heavier than Gaussian because sidelobes and unrelated sources fall on random positions.

\subsection{Forced detections and chance background} \label{sec:candidates}

A star is a forced detection if it reaches $S/\sigma\ge5$ in any epoch or in its inverse-variance-weighted coadd over epochs and is not a secure association with a natural explanation. Other secure associations, including the source near LSPM~J1943+5337 that is shown not to be at the star (Section~\ref{sec:lspm}), are tested like any other star, and the chance background must then count such coincidences: a component that would be a secure association of a control position ($R>0.5$ under the model of Section~\ref{sec:reliability}, evaluated at the Epoch-1 epoch) is exempt from that position's confusion veto, exactly as a star's own secure component is (\NCtrlAssoc\ control-position components). The control positions are measured in the epochs that survive the star's own veto, each with its own veto in those epochs, so that a star and its controls are compared over the same epochs. Cataloged Epoch-1 components near a star can put flux at its position through their beam wings, especially at low declination where the QL beam is elongated. We therefore model every component within 15\arcsec\ as an elliptical Gaussian with its fitted size and peak, and veto star-epochs where the predicted contribution reaches half the detection threshold; \NVetoStarEp\ star-epochs of \NStarsVeto\ stars are vetoed, and those stars receive confusion-limited limits.

For the steady test the predicted contribution is also coadded over epochs like the flux density, and a star is vetoed if it reaches half the coadd threshold (\NCoaddVeto\ stars). A genuine source at the star is not vetoed by its own component if that component is a secure association; under the reliability model a component at the star has $R<0.5$ with probability $2\pi\lambda\sigma_i^2/f$, about 1--3\% at catalog S/N\,=\,5--7, and none of the \NNatural\ natural emitters is vetoed. Forced detections are characterized with the same SIMBAD and X-ray evidence as the secure associations (Section~\ref{sec:classify}), but whether one counts as a possible transmitter depends on the eligibility of its star, defined below. We treat two kinds of transmitter separately, because they are best found with different statistics and therefore have different numbers of chance trials. A steady transmitter is best detected in the coadd, one trial per star; a transmitter that is on only intermittently is best detected in a single epoch, one trial per star-epoch. For each test we count the forced detections that would count as possible transmitters, $k$, whatever their origin, and estimate how many arise by chance, $b$, by applying the identical selection and veto at the control positions, where no stellar emission is expected. Given $k$ and $b$, the 95\% upper confidence limit $\mu_{95}$ on the number of genuine transmitter detections among them follows from $P(n\le k\mid\mu_{95}+b)/P(n\le k\mid b)=0.05$ \citep{Zech1989}; without background subtraction it is the classical Poisson upper limit for $k$, which we adopt for the headline limits (Section~\ref{sec:limits}).

A continuum detection cannot by itself distinguish a transmitter from stellar emission, so the search must specify at which stars, and above which luminosity, a detection would count as a possible transmitter. Excluding every star with activity evidence---an X-ray counterpart in the full 2RXS or eRASS1 catalog (each matched at the survey epoch, 1990.8 and 2020.3, with Gaia proper motions) or a SIMBAD activity type (Section~\ref{sec:classify})---is unsuitable, because that evidence measures how deep the X-ray surveys reach at the star's distance and how well the star has been studied, rather than its expected radio emission: \NActive\ stars (\PctActive\%) have it, but the fraction falls from \PctActTen\% within 10\,pc to \PctActHundred\% at 80--100\,pc, so such a rule would remove most of the nearest stars, which are the ones most sensitive to faint transmitters; even among the AFGK stars it would leave only \NOldEligAFGKTen\ of the \NAFGKTen\ within 10\,pc eligible (\PctOldEligAFGKTen\%), against \PctOldEligAFGKHundred\% at 80--100\,pc. We instead use criteria based on stellar type and on activity level, as follows:

\begin{enumerate}
\item M dwarfs ($G_{BP}-G_{RP}\ge1.84$) flare, and their gyrosynchrotron and coherent emission spans the luminosity range of our detections whether or not their activity is cataloged, so an M-dwarf detection cannot be recognized as artificial unless it exceeds any plausible natural emission. We adopt an empirical envelope $L_{\rm env}=10^{10}\,\WHz$ and count an M-dwarf detection only if its measured luminosity exceeds $L_{\rm env}$. No M-dwarf detection in any epoch of this survey reaches it: the most luminous, $\MaxMdwL\,\WHz$, is a single $\MaxMdwSNR\sigma$ measurement at a position where the noise is \MaxMdwSigRatio\ times the median, and the next most luminous is $\NextMdwL\,\WHz$. The luminous 15\,GHz incoherent flare observed from the young M-dwarf binary DG~CVn reached $\LDGCVn\,\WHz$ \citep{Fender2015}. Its spectrum was not measured, and the peak preceded the first measurement, but gyrosynchrotron flare spectra usually peak near 10\,GHz and are optically thick below the peak \citep{Osten2016}, so its luminosity at 3\,GHz was probably lower; an optically thin spectrum $S_\nu\propto\nu^{-0.5}$ down to 3\,GHz would raise it only to $\approx L_{\rm env}$. Rarer flares or coherent bursts, which are brightest at low frequencies, could exceed the envelope at 3\,GHz; such an event would be counted in $k$.

\item AFGK stars with Gaia evidence of multiplicity (RUWE\,$>1.4$, a non-single-star solution, or $\Delta M_G<-0.5$\,mag) may have an unresolved active or M-dwarf companion and are treated like M dwarfs. These are \NAFGKMult\ stars, \PctAFGKMult\% of the AFGK stars that would otherwise be eligible, a fraction nearly independent of distance (\PctMultAFGKTen\% within 10\,pc, \PctMultAFGKHundred\% at 80--100\,pc). RUWE is inflated for the brightest stars, of which \PctRuweBright\% at $G<6$ exceed 1.4 against \PctRuweFaint\% at $G=6$--13, so some, such as $\epsilon$~Eri (RUWE\,=\,\EpsEriRUWE), are probably flagged spuriously and are eligible only above $L_{\rm env}$.

\item Classes of known radio emitters are excluded at all luminosities: SIMBAD cataclysmic variables, RS~CVn binaries, young stars, and magnetic chemically peculiar stars, and Gaia DR3 eclipsing binaries \citep[\NEBall\ in the sample;][]{Mowlavi2023}, which include the Algol systems; the flares of RS~CVn and Algol binaries can exceed $10^{10}\,\WHz$ \citep{Gudel2002}. The most luminous stellar detection of any type in the survey, $\MaxStarL\,\WHz$ from the cataclysmic variable \MaxStarName\ in one epoch, is in these classes.

\item AFGK stars with saturated or nearly saturated coronae, $\log L_X/L_{\rm bol}\ge-4$, are also excluded at all luminosities (\NExclHot\ stars). $L_X$ is computed as in Section~\ref{sec:classify}, and $L_{\rm bol}$ from $M_G$ without a bolometric correction ($|BC_G|<0.5$\,mag for these stars, ample for a threshold at the 1-dex level). This criterion is a property of the star, but whether it can be applied is not uniform. We take nominal survey depths of $2\times10^{-13}$\,erg\,cm$^{-2}$\,s$^{-1}$ for 2RXS (0.1--2.4\,keV; \RxsDepthCR\,count\,s$^{-1}$ at the adopted conversion, close to the median 2RXS count rate of \RxsMedCR\,count\,s$^{-1}$, although the survey exposure varies across the sky) and $5\times10^{-14}$\,erg\,cm$^{-2}$\,s$^{-1}$ for eRASS1 (the hemisphere-median 0.5--2\,keV flux at 50\% completeness; \citealt{Merloni2024}), which we compare with the 0.2--2.3\,keV fluxes without a band correction. At these depths 2RXS reaches $\log L_X/L_{\rm bol}=-4$ at \XLimRxsPctFour\% of the AFGK stars and eRASS1, over half the sky, at \XLimEraPctFour\%, whereas a saturated corona, $\log L_X/L_{\rm bol}\approx-3$, is detectable at almost all of them ($\log L_X/L_{\rm bol}=\XLimRxsHi$ for 99\% with 2RXS). Table~\ref{tab:xray} shows, by distance, how many AFGK stars outside the known-emitter classes have a measured $\log L_X/L_{\rm bol}$ above or below the threshold, have no counterpart where the nominal depth reaches the threshold, or are unconstrained: the unconstrained fraction rises from zero within 20\,pc to \PctXUncHundred\% at 80--100\,pc. The eligible stars are therefore those without measured evidence of activity above the threshold, and an active star among the unconstrained ones would be counted in $k$. TYC~3573-1391-1 (\TycDist\,pc), the one eligible star that passes the steady test (Section~\ref{sec:cand_results}), is an example: it has no X-ray counterpart, and 2RXS reaches only $\log L_X/L_{\rm bol}\approx\TycXLim$ at its distance, so its activity is unconstrained at $-4$. A harder corona than assumed in the 2RXS conversion would raise $L_X$: for ${\rm HR1}=0$ or $+0.3$, \NEcfZero\ or \NEcfHard\ of the \NEligRxsOnly\ eligible stars whose $L_X$ rests on 2RXS alone would cross the threshold, none of them a counted detection.
\end{enumerate}

In all, \NElig\ AFGK stars (\PctElig\% of the sample) are eligible at any luminosity, \NEnv\ (\NEnvM\ M dwarfs and \NAFGKMult\ AFGK stars) only above $L_{\rm env}$, and \NExclAll\ at none.

The eligible fraction of AFGK stars is \PctEligAFGKTen\% within 10\,pc and \PctEligAFGKHundred\% at 80--100\,pc, nearly flat with distance, against \PctOldEligAFGKTen\% and \PctOldEligAFGKHundred\% under the presence rule. These rules err toward counting: natural emission at a star that we treat as eligible---from a multiple that Gaia does not flag, from a corona below the X-ray detection limit, or from an M-dwarf flare above $L_{\rm env}$---enters $k$, which weakens the limits rather than tightening them falsely. We adopted the multiplicity criterion after examining the forced detections at AFGK stars (Section~\ref{sec:cand_results}), and give the results without it. Activity evidence itself is used only to divide the stacks (Section~\ref{sec:stack_method}).

\begin{deluxetable}{lrrrrrr}
\tablecaption{X-ray constraints on the activity level of the AFGK stars outside the known radio-emitter classes.\label{tab:xray}}
\tablehead{\colhead{Shell} & \colhead{$N$} & \colhead{Measured} & \colhead{Measured} & \colhead{Depth} & \colhead{Unconstrained} & \colhead{Unc.} \\
\colhead{(pc)} & \colhead{} & \colhead{$\ge-4$} & \colhead{$<-4$} & \colhead{reaches $-4$} & \colhead{} & \colhead{(\%)}}
\startdata
0--10 & 43 & 1 & 35 & 7 & 0 & 0 \\
10--20 & 312 & 18 & 193 & 101 & 0 & 0 \\
20--40 & 2{,}702 & 123 & 944 & 1{,}489 & 146 & 5 \\
40--60 & 7{,}778 & 331 & 1{,}603 & 4{,}347 & 1{,}497 & 19 \\
60--80 & 14{,}303 & 520 & 1{,}784 & 7{,}307 & 4{,}692 & 33 \\
80--100 & 23{,}330 & 825 & 1{,}825 & 10{,}331 & 10{,}349 & 44 \\
\enddata
\tablecomments{Columns give the number of stars with a measured $\log L_X/L_{\rm bol}\ge-4$ (excluded) or $<-4$ (an X-ray counterpart below the threshold), with no counterpart where the nominal survey depth reaches the threshold, and with no counterpart where it does not (unconstrained). Nominal depths are $5\times10^{-14}$\,erg\,cm$^{-2}$\,s$^{-1}$ in the eRASS1 half of the sky (the hemisphere-median 0.5--2\,keV flux at 50\% completeness; \citealt{Merloni2024}), compared without a band correction with the 0.2--2.3\,keV fluxes, and $2\times10^{-13}$\,erg\,cm$^{-2}$\,s$^{-1}$ (0.1--2.4\,keV; 2RXS) elsewhere; survey exposure varies across the sky, so the depth columns are approximate. $L_{\rm bol}$ is from $M_G$. Only measured activity excludes a star: stars in the last three columns are eligible, so an active star among them would be counted in $k$.}
\end{deluxetable}

The chance background follows the same rules: we count control-position detections, and trials, around the eligible stars, and around the envelope stars only those whose flux density would exceed $L_{\rm env}$ at the star's distance, of which there are none. The envelope test is applied separately to the coadd and to each epoch, and also to the later epochs of the \NSecEnv\ envelope stars that are secure natural emitters in Epoch~1 (\NSecEnvEp\ star-epochs; the brightest reaches $\LSecEnvMax\,\WHz$), none of which exceeds $L_{\rm env}$. Over all stars the control positions would give $b=\BSteadyAll$ in the steady test and \BSnapAll\ in the single-epoch test; around the eligible stars $b=\BSteady$ and \BSnap. Because $b$ rests on \NCtrlSteady\ and \NCtrlSnap\ control detections, it is uncertain by $\pm\BSteadyErr$ and $\pm\BSnapErr$ (Poisson), or $\pm\BSteadyErrBoot$ and $\pm\BSnapErrBoot$ from a bootstrap over stars, which accounts for control detections clustered around the same star. Each of the eight control angles is also an independent realization of the null, directly comparable with $k$: in the steady test they give \CtrlAngleMin--\CtrlAngleMax\ detections. Marginalizing the limit over this uncertainty raises $\mu_{95}$ from \MuSteady\ to \MuSteadyMarg\ and from \MuSnap\ to \MuSnapMarg, and we quote the standard values. Control positions near the stars with emission could pick up sidelobes of that emission, but excluding them changes $b$ to \BSteadyNoEmit\ and \BSnapNoEmit. For the duty-cycle limit we also count stars rather than star-epochs: $k=\KHost$ eligible stars reach $5\sigma$ in at least one epoch, against $b=\BHost$ ($\mu_{95,{\rm host}}=\MuHost$).

\subsection{Limits on individual stars} \label{sec:limits_method}

For a star without a secure association we adopt as its limit the 95\%-completeness detection threshold
\begin{equation}
S_{\rm lim}=\frac{(5+1.645)\,\sigma_{\rm img}}{b_{\rm pk}},
\end{equation}
where $\sigma_{\rm img}$ is the mean island rms of the star's QL image, $b_{\rm pk}$ the QL peak bias of its half-epoch, and 5$\sigma$ is the PyBDSF detection threshold: a source at least this bright would be detected with 95\% probability. This is a statement about the sensitivity at the star, not a confidence bound conditioned on the measured flux density, and ``limit'' for a star without emission has this meaning throughout. If an unrelated component lies within 3\arcsec\ (\NConfused\ stars) the limit is raised to that component's flux density; for the \NDetected\ securely detected stars, the limit on any artificial component is instead the data-conditioned one-sided 95\% upper confidence bound, the measured flux density plus 1.645$\sigma$. The spectral luminosity limit is $\Lnu^{\rm lim}=4\pi d^2S_{\rm lim}$ and, for a transmitter whose power is spread uniformly over the 2--4\,GHz band, ${\rm EIRP}_{\rm lim}=\Lnu^{\rm lim}\times2$\,GHz. A transmitter of any bandwidth within 2--4\,GHz contributes its in-band power divided by 2\,GHz to the band-averaged flux density, so the EIRP limit applies to the total in-band power of a signal that is on during the observation, is not removed by radio-frequency-interference flagging, and is not resolved out or suppressed by the imaging. The flagging matters for narrow signals. The VLA calibration pipeline used for VLASS \citep{Lacy2020} automatically flags outliers in time and frequency. A signal confined to one 2\,MHz channel has 1000 times the band-averaged flux density in that channel: at our median limit, $\sim$0.5\,Jy, of order the noise of a single visibility, but a signal ten times brighter would be a strong outlier and is likely to be flagged. Our limits therefore apply to signals spread over many channels, and to signals confined to a few channels only near the detection threshold.

For the all-epoch analysis (Section~\ref{sec:forced}) we use the inverse-variance-weighted coadd over unvetoed epochs: the completeness threshold $S_{\rm lim}=(5+1.645)\,\sigma_{\rm coadd}$ for stars without emission and, where larger, the upper confidence bound $S_{\rm coadd}+1.645\,\sigma_{\rm coadd}$ for stars with emission; for the few stars whose every epoch is vetoed for confusion, the limit is the largest single-epoch $S+1.645\sigma$. A transmitter that is on during only one snapshot, at an unknown epoch, is excluded only if it would have been detected in whichever epoch it was on, so its limit is the single-epoch limit $(5+1.645)\,\sigma_e$ in the noisiest unvetoed epoch (or $S_e+1.645\,\sigma_e$, where larger, for stars with emission); in the lowest-noise epoch alone the limit would be lower. All limits refer to emission at the star's position. We do not account for the falloff in sensitivity in the forced photometry for a transmitter offset from the star, and as a result do not constrain a transmitter on a planet at a projected separation of $\gg1\arcsec\times(d/{\rm pc})$\,au.

\subsection{Population limits} \label{sec:pop_method}

If a fraction $f$ of stars host a steady transmitter of spectral luminosity $L_T$, which gives a flux density $S_T=L_T/4\pi d_i^2$ at star $i$, that star is detected with probability $p_i(L_T)=\Phi[(S_T-t_i)/\sigma_i]$, where $\Phi$ is the standard normal distribution function, $t_i=5\sigma_i$, the threshold that defines $k$, so that each detection, the candidates included, is counted as a possible transmitter. We do not attempt to separate a transmitter from natural emission at a detected star; treating the measured flux density as a natural baseline that a transmitter must exceed by $5\sigma_i$, i.e., $t_i=S_i+5\sigma_i$, would change the limits by at most \AddDiffPct\%. At the envelope stars $t_i=\max(5\sigma_i,S_{{\rm env},i})$ with $S_{{\rm env},i}=L_{\rm env}/4\pi d_i^2$, because a detection there counts only above $L_{\rm env}$: a transmitter with $L_T=L_{\rm env}$ is detected at an envelope star with probability $\approx0.5$, and these stars begin to contribute slightly below $L_{\rm env}$. The excluded stars, detected or not, have $p_i=0$; $\sigma_i$ and $S_i$ refer to the coadd. As a check that requires no decision about the origin of any detection, we also compute the limit for all M dwarfs with a coadd, with $t_i=5\sigma_i$ at every one and every coadd detection, natural emitters included, counted in $k$. Injection--recovery (Section~\ref{sec:forced}) validates this form.

The expected number of detected transmitters is $fN_{\rm eff}(L_T)$ with $N_{\rm eff}=\sum_ip_i$, so $f<\mu_{95}/N_{\rm eff}(L_T)$, with $\mu_{95}$ from the steady test. This is a 95\% upper confidence limit on $f$ in which each star's completeness enters through $p_i$. We adopt as the headline the background-free limit, in which $\mu_{95}$ is the classical Poisson limit for the $k$ steady-test detections, all treated as possible transmitters, and give the background-subtracted (Zech) limit for comparison: the background-free limit does not depend on the control-position calibration, and it is the more conservative of the two. For the Epoch-1 catalog analysis, which has no control-position background, this reduces to the classical Poisson limit $f<\mu_k/N_{\rm eff}$ for the $k=\KEOne$ unexplained secure associations ($\mu_k=\MuEOne$; with no background term, the source shown not to be at LSPM~J1943+5337 is excluded here), with $p_i$ a step function at each star's limit, the same eligibility rules, and the background source near LSPM~J1943+5337 counting only above its measured luminosity. A transmitter that is on during a random fraction $\delta$ of the few-second VLASS snapshots is detected at its host in at least one epoch with probability $1-\prod_e(1-\delta p_{ie})$, if it is on or off independently in each snapshot, where $p_{ie}$ are the single-epoch detection probabilities of an eligible star (for an envelope star, of a detection above $L_{\rm env}$). We compare the expected number of detected hosts, $f\sum_i[1-\prod_e(1-\delta p_{ie})]$, with $\mu_{95,{\rm host}}$ from the host-level single-epoch test (Section~\ref{sec:candidates}); because a host on in several epochs counts once, this avoids treating repeated detections of one star as independent. As $\delta\to0$ the expected number tends to $f\delta\sum_{i,e}p_{ie}$, giving $f\delta<\mu_{95,{\rm host}}/\sum_{i,e}p_{ie}$.

\subsection{Stacking} \label{sec:stack_method}

Image stacking of survey cutouts was developed for extragalactic samples \citep{White2007} and has since been applied to stars: to nearby young solar-type stars at 74\,MHz \citep{Lazio2010}, to nearby low-mass stars, exoplanet hosts, and young stellar objects at 0.07--1.4\,GHz \citep{Ling2022}, to the Taurus--Auriga members in VLASS \citep{Rahman2025}, and to cataclysmic variables \citep{Rao2026}. Here we stack every star of a volume-limited Gaia sample of field main-sequence stars, at its epoch-propagated position, by spectral type and activity. To probe emission below the single-star threshold we stack the forced photometry in six distance shells (0--10, 10--20, 20--40, 40--60, 60--80, and 80--100\,pc), each split at $G_{BP}-G_{RP}=1.84$ into AFGK stars (including \NAstars\ A stars bluer than 0.33\,mag) and M dwarfs, per epoch and for the all-epoch coadd. We form two statistics. The inverse-variance-weighted mean spectral luminosity
\begin{equation}
\langle\Lnu\rangle=\frac{\sum_i w_i\,4\pi d_i^2 S_i}{\sum_i w_i},\qquad w_i=\left(4\pi d_i^2\sigma_i\right)^{-2},
\end{equation}
uses only clean stars: no secure association, no forced detection, and no epoch with predicted catalog contamination above $0.5\sigma$. Its null is the same weighted mean at each of the eight control positions, which are subject to the same clean cuts as the stars; we subtract their average and adopt the larger of the formal error and their scatter. We report the stacks for all clean stars and separately for those with and without activity evidence (Section~\ref{sec:candidates}). Because the weights scale as $d^{-4}\sigma^{-2}$, we report the effective number of stars, $N_{\rm eff,w}=(\sum w)^2/\sum w^2$, and the fraction of the weight carried by the ten highest-weight stars. The second statistic is the median $S/\sigma$ of all unvetoed stars in a shell, including detected stars, relative to the median at the control positions, with bootstrap errors; it is insensitive to a few bright sources and does not require excluding detections.

For limits on transmitters we use only the eligible AFGK stars: the stacks probe luminosities far below $L_{\rm env}$, where no other star is an eligible host. A mean stack constrains the product of the transmitter fraction and power. For each $L_T$ we stack the clean stars at which a transmitter of that luminosity would be at most $3\sigma$ in the coadd, a selection fixed by the noise alone, so that a host is rarely removed from the stack by the clean cuts. If each star hosts such a transmitter with probability $f$, the weighted mean $m=\sum_ia_iL_i$ (with normalized weights $a_i$) has expectation $fL_T$ and, besides the measurement error $s_m$, a host-sampling variance $L_T^2f(1-f)/N_{\rm eff}$ with $N_{\rm eff}=1/\sum_ia_i^2$, which matters when few stars carry the weight. The 95\% limit $f_{95}$ solves $f_{95}L_T-q\,[s_m^2+L_T^2f_{95}(1-f_{95})/N_{\rm eff}]^{1/2}=\max(m,0)$, where $s_m$ is the formal error scaled by the rms of $S/\sigma$ at the control positions or, if larger, the scatter of the eight control-position means, in which case $q$ is the Student-$t$ quantile for seven degrees of freedom instead of 1.645. A median stack instead bounds $f$ itself once $S_T\gg\sigma$: by Monte Carlo, we find the fraction of the eligible stars, pooled over 0--100\,pc and so chosen before seeing the data, that must receive an extra $S_T/\sigma_i$ for the median to shift by more than the observed shift plus 1.645 times its error. Comparing the two distinguishes a stacked signal produced by many faint emitters from one produced by a few bright ones.

\section{Results} \label{sec:results}

\subsection{Radio-detected stars are natural emitters} \label{sec:matches}

\begin{figure}
\centering
\includegraphics[width=0.6\textwidth]{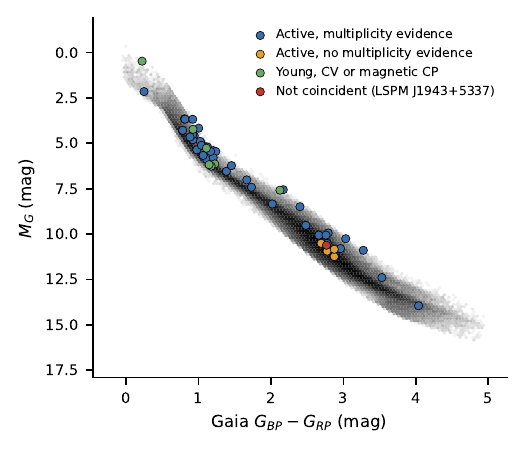}
\caption{Color--magnitude diagram of the \NSample\ sample stars (grey) with the \NSecure\ secure associations colored by evidence: active stars with multiplicity evidence (blue) and without it (orange), young stars, cataclysmic variables, and the magnetic chemically peculiar star (green), and the radio source near LSPM~J1943+5337, which is not coincident with the star (red).\label{fig:cmd}}
\end{figure}

Table~\ref{tab:matches} lists the \NSecure\ secure associations, and Figure~\ref{fig:cmd} shows them on the color--magnitude diagram. Of the \NCand\ stars with a VLASS component within 2\farcs5, \NSecure\ are secure associations. The other \NCandNonSec\ lie farther from their stars (median \asec{\NonSecSepMed}, against \asec{\SecSepMed} for the secure associations), and their number is close to the \NCandChance\ chance alignments expected from the offset catalogs; some are probably stellar emission that falls below the reliability threshold, such as V376~Cep (Section~\ref{sec:reliability}), and \NNonSecRxs\ of them have a 2RXS counterpart. Like every other star, they are tested by forced photometry in every epoch (Section~\ref{sec:forced}). Of all \NCand, \NRxs\ have a 2RXS and \NEra\ an eRASS1 counterpart; from the catalogs' source densities, the chance of an unrelated 2RXS source within 40\arcsec\ is $\sim\PChanceRxs$, and of an eRASS1 source within 20\arcsec\ $\sim\PChanceEra$.

Among the secure associations (Table~\ref{tab:matches}) are \NCFlare\ flare stars, \NCRS\ RS~CVn binaries, \NCBY\ BY~Dra variables, \NCYoung\ young stars, \NCCV\ cataclysmic variables (including AE~Aqr), and the magnetic Ap star $\alpha^2$~CVn; \NCXA\ are X-ray-active stars without a more specific SIMBAD type (\NXAMult\ of them with multiplicity evidence), and \NCEm\ is an emission-line star. \NSecureX\ of the \NNatural\ natural emitters have X-ray counterparts. Independently of class, \NGrpMult\ of the active stars show multiplicity evidence and \NGrpSingle\ do not; the remaining \NGrpOther\ are young stars, cataclysmic variables, or the chemically peculiar star. The X-ray-detected stars have $\log(L_X/\Lnu)=\GBmin$--\GBmax\ (median \GBmed), $\sim$1\,dex radio-bright relative to the G\"udel--Benz relation, as expected for a radio flux-limited selection and consistent with the radio-loud tail of active stars \citep{Vedantham2022}. The emission-line star, \EmSecName\ (an M dwarf at \EmSecDist\,pc; $R=\EmSecRel$, $S=\EmSecS$\,mJy), has multiplicity evidence (RUWE\,=\,\EmSecRUWE, \EmSecdMG\,mag over-luminous) but no X-ray counterpart; its activity evidence is its SIMBAD emission-line type. It was detected only in Epoch~1: forced photometry gives $S/\sigma=\EmSecSNRone$ in Epoch~1 and at most \EmSecSNRlater\ later, as expected for a flare. The remaining secure association is discussed in Section~\ref{sec:lspm}. Two further unexplained components within 1\arcsec\ of a star, near LP~732-30 and 2MASS~J14223872+0646143, have $R<0.01$ after the astrometric correction, and independently they did not move with their stars between NVSS or FIRST and VLASS (the stars moved 13\farcs5 and 1\farcs7, respectively): both are background sources.

Independent evidence confirms the stellar nature of most associations. \NSRSC\ are listed in the Sydney Radio Star Catalogue \citep{Driessen2024}, \NSRSCV\ of them with circularly polarized detections that by themselves establish stellar emission, and \NDeOverlap\ are in the VLASS--Gaia radio-star sample of \citet{De2024}. Of the \NDeTotal\ stars in that sample, \NDeInSample\ fall in ours. Those we do not securely associate are \NDeEpTwoOnly\ sources detected by \citet{De2024} only in VLASS Epoch~2, and Epoch-1 sources below the catalog threshold or with ambiguous matches ($R\le0.5$); at the latter our forced photometry agrees with their flux densities within $\sim$0.1\,mJy. Additionally, \NSRSCinSample\ Sydney Catalogue stars lie in our sample, and we securely associate \NSRSCdet\ of them. Of the other \NSRSCundet, \NSRSClow\ have cataloged detections only below 2\,GHz (mainly RACS and LoTSS), and \NSRSCvlass\ were detected in VLASS, \NSRSCvlassCand\ of which are forced detections (Section~\ref{sec:candidates}); in all, \NSRSCcand\ of the \NSRSCundet\ are forced detections, and the rest receive limits, as expected for variable emission. \citet{De2024} list LSPM~J1943+5337 as a confirmed radio star on the basis of its Epoch-1 coincidence; Section~\ref{sec:lspm} shows that the emission is not at the star.

{\setlength{\tabcolsep}{2.5pt}  
\startlongtable
\begin{deluxetable*}{lrrrrrrrcrccl}
\tabletypesize{\scriptsize}
\tablecaption{VLASS Epoch 1 components securely associated (reliability $R>0.5$) with \NSecure\ main-sequence stars within 100\,pc.\label{tab:matches}}
\tablehead{\colhead{Name} & \colhead{Gaia DR3} & \colhead{$d$} & \colhead{$G_{BP}-G_{RP}$} & \colhead{Sep.} & \colhead{$R$} & \colhead{$S_{3\,\rm GHz}$} & \colhead{$\log L_\nu$} & \colhead{X-ray} & \colhead{$\log\frac{L_X}{L_\nu}$} & \colhead{Mult.} & \colhead{Radio} & \colhead{Class} \\
\colhead{} & \colhead{} & \colhead{(pc)} & \colhead{(mag)} & \colhead{(arcsec)} & \colhead{} & \colhead{(mJy)} & \colhead{(W\,Hz$^{-1}$)} & \colhead{} & \colhead{(Hz)} & \colhead{} & \colhead{} & \colhead{}}
\startdata
{G 9-38B} & 660597997697274752 & 5.1 & 4.04 & 0.69 & 0.920 & 4.34 & 7.13 & E & 13.29 & SR & Pc & Fl \\
{BD+19 5116A} & 2824770686019003904 & 6.3 & 2.66 & 0.22 & 0.983 & 1.13 & 6.72 & R & 15.14 & S & PDvc & Fl \\
{V374 Peg} & 1893189736896880640 & 9.1 & 2.87 & 0.57 & 0.862 & 1.48 & 7.17 & R & 14.24 & \nodata & KDvc & Fl \\
{CW UMa} & 758200712885953280 & 13.4 & 2.69 & 0.12 & 0.986 & 1.99 & 7.63 & RE & 13.95 & \nodata & Kvc & Fl \\
{alf For B} & 5059349158319689344 & 14.1 & 1.11 & 0.10 & 0.998 & 5.04 & 8.08 & RE & 14.46 & SR & Pvc & X \\
{CD-38 11343B} & 5970189504209690496 & 15.7 & 2.96 & 1.11 & 0.654 & 1.94 & 7.76 & RE & 15.04 & SO & PD & X \\
{1RXS J055641.0-101837} & 2999254213453678720 & 15.9 & 2.78 & 0.27 & 0.984 & 1.01 & 7.49 & RE & 13.61 & \nodata & Dv & X \\
{G 249-36} & 1005873614080407296 & 16.3 & 2.87 & 0.06 & 0.999 & 9.36 & 8.47 & R & 12.94 & \nodata & KDvc & X \\
{HD 43162C} & 2913314288686472576 & 16.7 & 2.78 & 0.10 & 0.997 & 3.28 & 8.04 & E & 13.95 & SR & KDv & X \\
{HD 283750} & 151650076838458112 & 17.4 & 1.38 & 0.09 & 0.998 & 2.96 & 8.03 & R & 14.74 & SNR & PDvc & BY \\
{LQ Hya} & 5739942011185493888 & 18.3 & 1.14 & 0.21 & 0.989 & 1.32 & 7.72 & RE & 15.07 & \nodata & vc & Y \\
{Ross 776} & 1851868127125091840 & 19.7 & 2.79 & 0.02 & 0.998 & 2.99 & 8.14 & R & 13.93 & SRO & PDv & Fl \\
{BD+55 1823} & 1429437378934511616 & 20.1 & 2.17 & 0.23 & 0.988 & 1.79 & 7.94 & R & 14.51 & SRO & PDv & Fl \\
{sig CrB A} & 1328866562170960512 & 22.7 & 0.81 & 0.18 & 0.991 & 4.00 & 8.39 & R & 15.15 & SO & KDvc & X \\
{BF Lyn} & 812477352773908608 & 23.4 & 1.24 & 0.21 & 0.991 & 1.04 & 7.83 & RE & 15.24 & SNO & Pvc & BY \\
{[R78b] 233} & 3274291270411229056 & 26.8 & 3.03 & 0.06 & 0.995 & 1.85 & 8.20 & RE & 13.70 & RO & KDvc & Fl \\
{V478 Lyr} & 2042389925739687808 & 27.1 & 0.98 & 0.12 & 0.997 & 2.74 & 8.38 & R & 14.67 & S & KDvc & RS \\
{StKM 1-497} & 3205094369407459456 & 27.8 & 2.12 & 0.29 & 0.982 & 1.58 & 8.16 & RE & 14.27 & SRO & Dv & Y \\
{PW And} & 2860839993168687872 & 28.3 & 1.17 & 0.17 & 0.995 & 1.72 & 8.21 & R & 14.73 & \nodata & Dv & Y \\
{HD 124498A} & 6298167775459045632 & 29.0 & 1.73 & 0.02 & 0.984 & 0.89 & 7.95 & RE & 14.62 & SR & PDvc & X \\
{2MASS J17022950+1341015} & 4544627976777952640 & 30.5 & 3.53 & 0.93 & 0.767 & 1.35 & 8.18 & \nodata & \nodata & RO & v & Em \\
{alf02 CVn} & 1517698716348324992 & 30.6 & 0.22 & 0.34 & 0.970 & 1.71 & 8.28 & R & 13.54 & SRO & KDvc & CP \\
{HD 166181} & 4590455518346609792 & 32.1 & 0.92 & 0.12 & 0.995 & 1.01 & 8.09 & R & 15.28 & SR & KDv & RS \\
{GS Leo} & 589720969909457792 & 35.4 & 1.06 & 0.20 & 0.992 & 1.87 & 8.45 & RE & 14.36 & SNR & PDvc & BY \\
{IZ Boo} & 1484942233437321728 & 36.0 & 2.40 & 0.21 & 0.992 & 2.00 & 8.49 & R & 13.73 & O & KDv & Fl \\
{HD 22403} & 69443337640657536 & 36.6 & 0.92 & 0.15 & 0.996 & 2.27 & 8.56 & R & 15.04 & SO & KDvc & Y \\
{NQ Com} & 3955187043030098176 & 40.5 & 1.67 & 0.36 & 0.977 & 0.86 & 8.23 & RE & 14.44 & RO & v & BY \\
{HD 143937} & 6243343170769568128 & 42.9 & 1.10 & 0.16 & 0.994 & 2.03 & 8.65 & RE & 14.59 & SO & KDv & X \\
{HD 195434A} & 4245659976227848704 & 45.1 & 1.20 & 0.17 & 0.990 & 1.32 & 8.51 & R & 14.70 & S & KD & X \\
{V402 Hya} & 5756823564825367936 & 47.5 & 1.17 & 0.26 & 0.987 & 2.64 & 8.85 & RE & 14.41 & RO & KDv & BY \\
Gaia DR3 3821318547411001728 & 3821318547411001728 & 49.0 & 2.48 & 0.22 & 0.991 & 1.33 & 8.58 & RE & 13.51 & R & Dv & X \\
{HD 200391} & 1845206534070618624 & 50.5 & 0.81 & 0.50 & 0.957 & 2.48 & 8.88 & R & 14.79 & SNO & Dv & RS \\
{HD 25284} & 4888246251178417024 & 50.6 & 1.45 & 0.20 & 0.990 & 1.72 & 8.72 & RE & 14.37 & SRO & D & X \\
{BO Mic} & 6778413151435607680 & 51.0 & 1.20 & 0.20 & 0.993 & 3.32 & 9.02 & R & 15.10 & SNRO & KDvc & BY \\
{EV Dra} & 1403766939225964288 & 57.5 & 0.94 & 0.06 & 0.998 & 1.74 & 8.84 & R & 14.50 & SNO & KDv & RS \\
{UCAC4 747-046156} & 860453786736413568 & 57.7 & 3.28 & 0.39 & 0.965 & 1.96 & 8.89 & R & 12.86 & O & Dvc & X \\
{HD 79551} & 818854107897169024 & 60.1 & 0.88 & 0.46 & 0.901 & 1.13 & 8.69 & R & 14.67 & S & Dv & X \\
{BD+47 4087} & 1942527533757602176 & 60.3 & 1.04 & 0.25 & 0.988 & 1.05 & 8.66 & R & 14.53 & SO & Dv & X \\
{HD 295290} & 3105605708307826944 & 60.5 & 1.02 & 0.12 & 0.997 & 6.65 & 9.46 & RE & 13.94 & O & Dv & Fl \\
{BD+18 1512} & 3361628369521591808 & 61.7 & 1.03 & 0.11 & 0.997 & 9.13 & 9.62 & RE & 13.75 & RO & Dv & X \\
{HD 8358} & 2534358706998049792 & 62.3 & 1.00 & 0.06 & 0.999 & 4.29 & 9.30 & R & 14.33 & SO & PDvc & RS \\
{RZ Cas} & 541801332594262912 & 65.3 & 0.25 & 0.07 & 0.999 & 3.25 & 9.22 & R & 14.13 & SN & Dv & X \\
{XY UMa} & 1023682919309621632 & 67.9 & 1.12 & 0.26 & 0.988 & 1.79 & 8.99 & R & 14.08 & SO & Dvc & RS \\
{1RXS J204412.7+312642} & 1859797431337296384 & 68.5 & 2.02 & 0.20 & 0.986 & 2.75 & 9.19 & R & 13.09 & R & KD & X \\
Gaia DR3 4808527947311614208 & 4808527947311614208 & 69.5 & 2.76 & 0.65 & 0.902 & 0.91 & 8.72 & RE & 13.14 & RO & v & X \\
{HD 101207B} & 767786259553138816 & 81.1 & 0.92 & 0.10 & 0.990 & 0.72 & 8.75 & R & 14.58 & SRO & \nodata & X \\
{BD+08 102} & 2750199780197751936 & 81.1 & 1.11 & 0.10 & 0.997 & 2.01 & 9.20 & R & 14.14 & SNRO & Kv & CV \\
{BD+66 1664} & 2210120493272273920 & 89.4 & 0.92 & 0.45 & 0.925 & 1.86 & 9.25 & R & 14.42 & SO & Dv & X \\
{AE Aqr} & 4226332451596335616 & 91.6 & 1.22 & 0.05 & 0.996 & 2.87 & 9.46 & R & 14.29 & S & KDvc & CV \\
{HD 156848B} & 4108078979665915648 & 93.4 & 0.78 & 0.28 & 0.986 & 1.29 & 9.13 & RE & 14.32 & S & Dv & X \\
{LSPM J1943+5337} & 2137474969751117312 & 97.6 & 2.77 & 0.26 & 0.987 & 1.81 & 9.31 & \nodata & \nodata & \nodata & D & BG \\
\enddata
\tablecomments{Sorted by distance. Separation at the epoch of the component's VLASS image. $S_{3\,\rm GHz}$ is the cataloged total flux density corrected for the Quick Look integrated-flux bias (divided by 0.90 in half-epoch 1.1; \citealt{Kimball2026}). X-ray: R = ROSAT 2RXS, E = eRASS1. Multiplicity evidence (Mult.): S = SIMBAD binary type, N = Gaia non-single-star solution, R = RUWE\,$>1.4$, O = over-luminous by $>0.5$\,mag. Radio-star evidence: P = in the Sydney Radio Star Catalogue \citep{Driessen2024} with a circularly polarized detection, K = in that catalog without one, D = in \citet{De2024}, v = significantly variable and c = co-moving with the star across VLASS epochs 1.1--4.1 (Section~\ref{sec:multiepoch}). Class (SIMBAD object types, except LQ~Hya and PW~And, young single stars; \citealt{Lehtinen2012,LopezSantiago2003}): Fl = flare star; Em = emission-line star; RS = RS~CVn binary; BY = BY~Dra variable; Y = young star; CV = cataclysmic variable; X = X-ray-active star; CP = magnetic chemically peculiar star; BG = not coincident with the star (Section~\ref{sec:lspm}). Machine-readable version in the reproducibility package.}
\end{deluxetable*}
}

\subsection{LSPM J1943+5337: emission not coincident with the star} \label{sec:lspm}

\begin{figure*}
\centering
\includegraphics[width=0.95\textwidth]{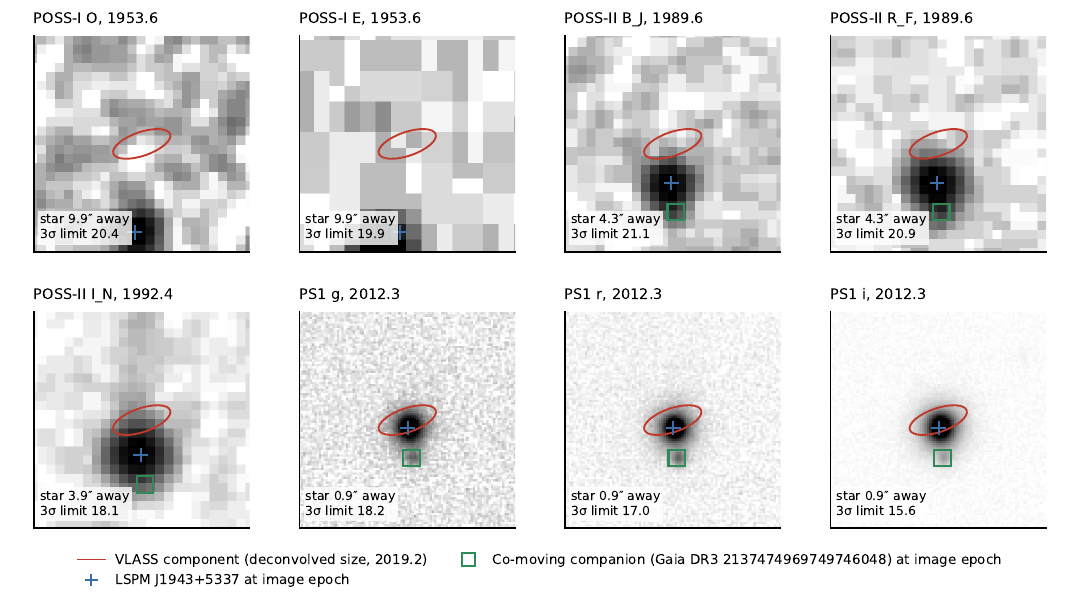}
\caption{Archival optical images ($24\arcsec\times24\arcsec$, north up, east left) around the VLASS component near LSPM~J1943+5337 (red ellipse: deconvolved size, at its Epoch-1 position, 2019.2). The blue cross marks the star and the green square its co-moving companion at each image epoch, from Gaia~DR3; their propagated $1\sigma$ positional uncertainties are at most \GaiaErrStar\ and \GaiaErrComp\,mas respectively, far smaller than the symbols. The star was displaced by \asec{\LspmOffPossI} (POSS-I) and $\sim4\arcsec$ (POSS-II) from the radio position; 3$\sigma$ limits at the radio position are given in each panel. At the Pan-STARRS1 epoch the star is only \asec{\LspmOffPS} away and dominates the light. Because the radio position is from 2019.2, this figure alone tests only for an optical counterpart of a stationary radio source; the multi-epoch astrometry of Figure~\ref{fig:lspm_vlass} shows that the radio source is not at the star's position, although it cannot yet exclude motion with the star at a fixed offset (Section~\ref{sec:lspm}).\label{fig:lspm_optical}}
\end{figure*}

The radio component \asec{\LspmSep} from the M dwarf LSPM~J1943+5337 ($d=\LspmDist$\,pc; $R=\LspmRel$) has a 3\,GHz flux density of \LspmS\,mJy and no X-ray counterpart, and the star has no SIMBAD activity classification; it is the only secure association without a natural explanation. Its cataloged deconvolved size, $\asec{\LspmMaj}\times\asec{\LspmMin}$, suggests a resolved, extragalactic source, although Quick Look sizes are uncertain at this signal-to-noise ratio \citep{Gordon2021}. The star moves north at 153\,mas\,yr$^{-1}$, so in POSS-I (1953) it lay \asec{\LspmOffPossI} and in POSS-II (1989--1992) $\sim4\arcsec$ from the radio position (under the assumption that the radio source did not move; see below). No optical source is present at the radio position on these plates to $g\gtrsim\PossIO$ and $r\gtrsim\PossIE$ (POSS-I) and $B_J\gtrsim\PossIIB$ and $R_F\gtrsim\PossIIR$ (POSS-II; Figure~\ref{fig:lspm_optical}). Because only $\sim$30\% of mJy radio sources have optical counterparts even at $r\sim22$ \citep{Ivezic2002}, and fewer at plate depth \citep{deVries2004}, this non-detection is weak evidence on its own. The archival imaging also reveals a common-proper-motion companion, Gaia~DR3~2137474969749746048, 3\farcs4 south of the star, whose faintness and colors suggest a cool white dwarf.

\begin{figure*}
\centering
\includegraphics[width=0.95\textwidth]{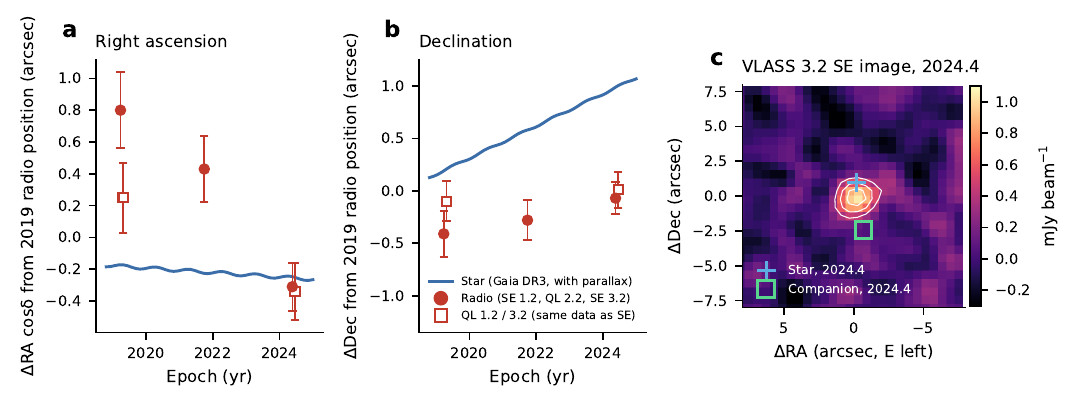}
\caption{(a) Right-ascension and (b) declination offsets of the radio source near LSPM~J1943+5337 relative to its Epoch 1 catalog position (red), in a frame tied to compact reference sources, compared with the Gaia track of the star including parallax (blue). Filled symbols are independent epochs; open squares are Quick Look images sharing data with the Single Epoch images. (c) VLASS 3.2 Single Epoch image (2024.4) with contours at 3, 5, and 7 times the rms; the star (cross) has moved 1\farcs1 from the radio source; the square marks its companion, as in Figure~\ref{fig:lspm_optical}.\label{fig:lspm_vlass}}
\end{figure*}

The decisive test is astrometric. Table~\ref{tab:lspm} and Figure~\ref{fig:lspm_vlass} show the radio position in VLASS epochs 1.2, 2.2, and 3.2, after tying each image to five compact reference sources; the reference frame agrees with Gaia to within $(\Delta\alpha^*,\Delta\delta)=(\asec{\FrameTieE},\asec{\FrameTieN})$ for the secure associations. The source is detected at a steady $\sim$1\,mJy\,beam$^{-1}$ in every epoch while the star moved $\sim$0\farcs8 north. In declination the radio position is constant ($\chi^2=\LspmChiDecStat$ for two degrees of freedom about its mean), and by 2024.4 it lies $\asec{\LspmOffThree}$ from the star (\LspmSigThree$\sigma$ in declination). Emission at the star's position is excluded with $\chi^2=\LspmChi$ for three degrees of freedom in declination and $\chi^2=\LspmChiTwoD$ for \LspmDofTwoD\ in both coordinates. In right ascension, along which the star barely moves, the radio positions scatter by more than their formal errors (a stationary and a co-moving source give $\chi^2=\LspmChiRAstat$ and \LspmChiRAcomove\ for two degrees of freedom). The source's deconvolved major axis lies at position angle \LspmPA$^\circ$, close to east--west, and QL and Single Epoch images made from the same 2019 visibilities differ by \asec{\LspmQLSEdE} in right ascension and \asec{\LspmQLSEdN} in declination. We therefore attribute the scatter to imaging-dependent centroid shifts of a slightly resolved source; inflating the right-ascension errors by a factor of three, enough to absorb it, leaves the declination test unchanged.

The radio emission is therefore not coincident with LSPM~J1943+5337. The present three-epoch baseline does not decide whether the source is stationary or shares the star's motion at a fixed offset: the fitted declination motion, $\LspmMu\pm\LspmMuErr$\,mas\,yr$^{-1}$, lies between zero and the star's 153\,mas\,yr$^{-1}$, and the all-epoch analysis of Section~\ref{sec:multiepoch} is also inconclusive. Positions measured directly in the Quick Look cutouts of the three epochs (Section~\ref{sec:cand_results}) favor a stationary source ($\chi^2=\LspmQLChiStat$ against \LspmQLChiCo\ for a co-moving one, for \LspmQLDof\ degrees of freedom), but QL positions of this slightly resolved source depend on the imaging, which is why Table~\ref{tab:lspm} uses Single Epoch positions where they exist; we do not regard the QL result as decisive. A co-moving source would lie $\sim\asec{\LspmOffArcsec}$ ($\sim\LspmOffAU$\,au projected) from the star, with $\Lnu\approx\LspmL\,\WHz$. Its extended structure, the absence of an X-ray counterpart, and the \NOffOne\ chance alignments within 1\arcsec\ expected across the sample (Section~\ref{sec:xmatch}) favor a background source, as does the reliability analysis, which predicts \NFalseSecure\ chance alignments among the secure associations (Section~\ref{sec:reliability}); we classify it as a likely background source. The chance background in the all-epoch tests includes such coincidences, so the source would be counted there if the star were eligible; as an M dwarf below $L_{\rm env}$ it is not (Section~\ref{sec:candidates}). It is nonetheless the one association in the sample whose nature is not settled. A VLBI detection of a compact component at the star's position would show the emission to be stellar; a non-detection would be less informative, because an arcsecond-scale source can be resolved out.

\begin{deluxetable}{lrrrrrrr}
\tablecaption{Multi-epoch VLASS astrometry of the radio source near LSPM~J1943+5337.\label{tab:lspm}}
\tablehead{\colhead{Image} & \colhead{Epoch} & \colhead{$S_{\rm peak}$} & \colhead{SNR} & \colhead{$\Delta\alpha^*$} & \colhead{$\Delta\delta$} & \colhead{$\Delta\alpha^*_\star$} & \colhead{$\Delta\delta_\star$} \\
\colhead{} & \colhead{(yr)} & \colhead{(mJy\,b$^{-1}$)} & \colhead{} & \colhead{(arcsec)} & \colhead{(arcsec)} & \colhead{(arcsec)} & \colhead{(arcsec)}}
\startdata
QL 1.2 & 2019.23 & 0.92 & 7.5 & $+0.25\pm0.22$ & $-0.10\pm0.19$ & $-0.17$ & $+0.19$ \\
QL 2.2 & 2021.74 & 1.03 & 7.8 & $+0.43\pm0.21$ & $-0.28\pm0.19$ & $-0.23$ & $+0.58$ \\
QL 3.2 & 2024.38 & 1.04 & 8.3 & $-0.34\pm0.18$ & $+0.01\pm0.17$ & $-0.25$ & $+0.99$ \\
SE 1.2 & 2019.23 & 0.78 & 6.2 & $+0.80\pm0.24$ & $-0.41\pm0.22$ & $-0.17$ & $+0.19$ \\
SE 3.2 & 2024.38 & 1.06 & 8.5 & $-0.31\pm0.15$ & $-0.07\pm0.15$ & $-0.25$ & $+0.99$ \\
\enddata
\tablecomments{Offsets are relative to the Epoch 1 catalog position, after tying each image to five compact reference components. Subscript $\star$: Gaia position of the star at the image epoch.}
\end{deluxetable}

\subsection{Multi-epoch behavior of the detected stars} \label{sec:multiepoch}

To test whether the secure associations behave as stellar emitters, we measured each one in every VLASS QL image covering it, from epoch 1.1 (2017) to 4.1 (2025--2026). For each image we extracted $61\arcsec\times61\arcsec$ cutouts around the star and around up to six compact, isolated reference components within $0\fdg45$ (Epoch-1 catalog peak flux density $>4$\,mJy\,beam$^{-1}$, deconvolved major axis $<1\farcs5$, nearest neighbor $>30\arcsec$). The references fix each image's astrometric frame relative to the corrected Epoch-1 positions and measure its flux scale relative to the Epoch-1 catalog (\FluxScaleMin--\FluxScaleMax\ per half-epoch, used only as a check; flux densities are divided by $b_{\rm pk}$). At the star we fitted a two-dimensional Gaussian if a $\ge5\sigma$ peak lay within 3\arcsec\ of the propagated position, and otherwise recorded the forced peak value, with $\sigma$ the robust rms of the cutout beyond 10\arcsec. Position errors combine the fit error, the frame-tie error, and a 0\farcs1 systematic floor in quadrature. Keeping one image per star per half-epoch gives \NMultiImg\ independent measurements (\NImgPerStarMin--\NImgPerStarMax\ per star), of which \NMultiDetImg\ are detections.

\begin{figure*}
\centering
\includegraphics[width=0.95\textwidth]{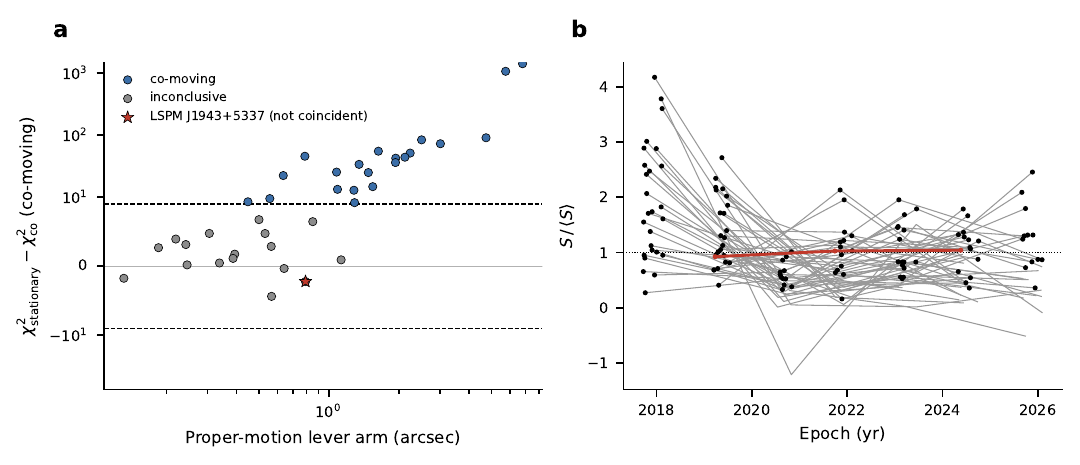}
\caption{Multi-epoch behavior of the \NSecure\ secure associations in VLASS epochs 1.1--4.1. (a) $\chi^2$ difference between a stationary source and one moving with the star at a constant offset, against the star's proper-motion lever arm over the detected epochs; dashed lines mark $|\Delta\chi^2|=9$. Blue: co-moving; grey: inconclusive; red star: LSPM~J1943+5337. Sources detected in only one epoch are not shown. (b) Peak flux density relative to each source's mean, by epoch (dots: detections). LSPM~J1943+5337 (red) is steady; the excess at Epoch~1 reflects selection.\label{fig:multiepoch}}
\end{figure*}

Figure~\ref{fig:multiepoch} summarizes the results. \NVariable\ of the \NSecure\ sources are variable ($p<0.01$ for a constant flux density), as expected for flaring and gyrosynchrotron emitters; because the sample was selected in Epoch~1, the light curves are biased high at that epoch. For the positions we compare a source on the star's track, allowing a constant offset, with a stationary source. \NComoving\ sources are co-moving ($\Delta\chi^2>9$; proper-motion lever arms \asec{\LeverMin}--\asec{\LeverMax}, median \asec{\LeverMed}), \NInconclusive\ are inconclusive because the lever arm is short compared with the position errors, and \NSingleDet\ are detected in only one epoch; none prefers a stationary source. Every source with a sufficient lever arm therefore moves with its star. Two co-moving or inconclusive sources, G~9-38B and CD-38~11343B, lie \asec{\GNineOff} and \asec{\CDOff} from the Gaia position of the matched star; both are members of multiple systems, and the emission may arise in a companion.

\subsection{Forced detections in all epochs} \label{sec:cand_results}

Forced photometry yields \NCandStars\ forced detections (\NCandCoadd\ in the coadd and \NCandEpoch\ in at least one epoch; Figure~\ref{fig:limits}). \NCandM\ are M dwarfs, all below $L_{\rm env}$ (the most luminous reaches $\MaxForcedML\,\WHz$ in its brightest epoch), and \NCandAFGKExcl\ are AFGK stars of the excluded classes or with $\log L_X/L_{\rm bol}\ge-4$, so none of these counts as a possible transmitter. Most are known natural emitters: \NCandActEv\ have activity evidence, flare stars and active binaries detected in a single epoch or persistent emitters below the Epoch-1 catalog threshold. Of these, \NCandXcat\ have a 2RXS or eRASS1 counterpart, and the other \NCandSimbadOnly\ are identified by SIMBAD object types alone; SIMBAD lists \NCandTX\ as X-ray sources, \NCandTFl\ as flare stars, \NCandTEm\ as emission-line stars, \NCandTBY\ as BY~Dra and \NCandTRS\ as RS~CVn variables, \NCandTY\ as young stars, and \NCandTCP\ as a magnetic chemically peculiar star (a star can have several types).

Table~\ref{tab:unexplained} lists the \NAFGKCand\ forced detections at AFGK stars outside the excluded classes. \NAFGKCandMult\ of them are at stars with Gaia multiplicity evidence, including the two most significant, 12~Sco~B and V335~Peg (16.6$\sigma$ and 16.7$\sigma$ in a single epoch; V335~Peg is an eclipsing and spectroscopic binary in SIMBAD that is not in the Gaia DR3 eclipsing-binary table), and none reaches $L_{\rm env}$. If these stars were eligible at all luminosities, the AFGK stars would give $k=\KSteadyWithMult$ against $b=\BSteadyWithMult$ in the steady test ($P(n\ge k\mid b)=\PSteadyWithMult$) and $k=\KSnapWithMult$ against $b=\BSnapWithMult$ in the single-epoch test ($P=\PSnapWithMult$). This significant excess, \NAFGKCandMult\ of \NAFGKCand\ detections at the \PctAFGKMult\% of stars with multiplicity evidence, is what unrecognized active or M-dwarf companions would produce, given that close binaries dominate the radio-detected stars (Section~\ref{sec:matches}); it led us to adopt the multiplicity criterion of Section~\ref{sec:candidates}, and it would be over-interpreted as a population of transmitters.

The remaining \KUnexpl\ forced detections, at eligible stars, are counted in $k$. Only TYC~3573-1391-1 passes the steady test ($\TycSNR\sigma$ in the coadd), and its emission is offset from the star: the one forced detection bright enough for a position in \CandAstroNdet\ epochs, it is fitted at $\CandAstroS$\,mJy $\asec{\CandAstroOff}$ from the star, against $\CandAstroScoadd$\,mJy measured at the star, and emission at the star's position is excluded ($\chi^2=\CandAstroChi$ for \CandAstroDof\ degrees of freedom); four independent catalogs place it 1\farcs7--1\farcs8 west of the star (Section~\ref{sec:catcheck}). Gaia DR3 lists no source at the radio position, and the star moves only \asec{\TycMotion} over the four epochs, too little to tell a stationary from a co-moving source ($p=\TycPStat$ and \TycPCo). SIMBAD lists no object within \TycFieldSimR\arcmin\ of the radio position other than the star (the nearest is an unrelated Gaia star \asec{\TycFieldSimNear} away), and Gaia DR3 no other source within \asec{\TycFieldGaiaNear}; \TycNvssName\ ($\TycNvssS\pm\TycNvssES$\,mJy at 1.4\,GHz), \asec{\TycNvssSep} away and within $\TycNvssSig\sigma$ of its position errors, is probably the same source. The emission is therefore either a background source, which the control positions at the same Galactic latitude ($b=\TycFieldB^\circ$) account for, or a faint companion $\sim$\TycProjAU\,au from the star in projection, which Gaia might not detect so close to a $G=\TycG$ star; either way it is counted in $k$. The other two, 1~Boo and TYC~7253-259-1, reach $\SnapOnlyLo$--$\SnapOnlyHi\sigma$ in a single epoch only, as expected for noise peaks or for flares from undetected companions, but also for a transmitter with a low duty cycle; 1~Boo is a double star in SIMBAD without Gaia multiplicity evidence, an example of the residual multiples that remain among the eligible stars.

In the steady test $k=\KSteady$ compares with $b=\BSteady$, giving $\mu_{95}=\MuSteady$ ($P(n\ge k\mid b)=\PSteadyHigh$), and in the single-epoch test $k=\KSnap$ star-epochs compare with $b=\BSnap$, giving $\mu_{95,{\rm snap}}=\MuSnap$ ($P=\PSnapHigh$): at the eligible stars there is no excess of forced detections over chance coincidences with background sources and noise. The eight control angles give \CtrlAngleMin--\CtrlAngleMax\ steady-test detections (rank probability \PRankSteady). The agreement holds at other thresholds: at $\ge4\sigma$ the eligible stars have \KCoFour\ coadd detections against \BCoFour\ expected, at $\ge7\sigma$ \KCoSeven\ against \BCoSeven, and the negative tails agree (\KCoNegThree\ against \BCoNegThree\ below $-3\sigma$). Around the same stars the Epoch-1 catalog has \CatStarIn\ components at 1\arcsec--2\farcs5 from the stars, against \CatCtrlIn\ around control points 14\arcsec--3\arcmin\ away.

The M-dwarf forced detections include steady emitters without activity evidence under our criteria, which illustrate why an M dwarf cannot be searched individually by its continuum flux. G~133-48 (2MASS~J01534955+4427284), a \GOTEd\,pc M dwarf, has a steady $\Lnu\approx\GOTEL\,\WHz$, within the range of the \NMSecNat\ natural-emitter M~dwarfs among our secure associations ($\MSecLlo$--$\MSecLhi\,\WHz$). It is detected \asec{\GOTEsep} from the star in an independent catalog (Section~\ref{sec:catcheck}). We fitted an elliptical Gaussian with the restoring beam of each image (from its header) to the Quick Look cutout of each of its three epochs (fitted S/N\,=\,\GoteSnrLo--\GoteSnrHi), calibrating the position errors by re-fitting the same source injected into \NInjAstro\ other cutouts of that epoch; the same procedure applied to LSPM~J1943+5337 recovers its offset from the star (Section~\ref{sec:lspm}). Emission at the star fits well ($\chi^2=\GoteChiStar$ for \GoteDofStar\ degrees of freedom), whereas a stationary source fits less well over the star's \asec{\GoteMotion} motion ($\chi^2=\GoteChiStat$ for \GoteDofStat, $p=\GotePStat$). The star has no X-ray counterpart or SIMBAD activity type, but it is a rapid rotator, with a photometric period of 0.216\,d \citep{Newton2016} and $v\sin i=47.6\pm2.9$\,km\,s$^{-1}$ \citep{Kesseli2018}, and it is over-luminous by \GOTEdMG\,mag, which suggests an unresolved companion: magnetic stellar emission is the natural explanation, but nothing in the continuum flux distinguishes it from a transmitter of the same luminosity. LSPM~J1943+5337 is the likely background source of Section~\ref{sec:lspm}.

{\setlength{\tabcolsep}{3pt}
\begin{deluxetable*}{lrcrrrr>{\raggedright\arraybackslash}p{2.5cm}r>{\raggedright\arraybackslash}p{2.4cm}>{\raggedright\arraybackslash}p{2.0cm}}
\tabletypesize{\scriptsize}
\tablecaption{Forced detections (any epoch or the coadd at $\ge5\sigma$) at AFGK stars outside the known radio-emitter classes and without measured evidence for $\log L_X/L_{\rm bol}\ge-4$.\label{tab:unexplained}}
\tablehead{\colhead{Name} & \colhead{Gaia DR3} & \colhead{$k$} & \colhead{$d$} & \colhead{$G_{BP}-G_{RP}$} & \colhead{$S_{\rm coadd}$} & \colhead{SNR} & \colhead{Half-epoch:\,SNR} & \colhead{$L_{\nu,\max}$} & \colhead{Multiplicity} & \colhead{Catalogs} \\
\colhead{} & \colhead{} & \colhead{} & \colhead{(pc)} & \colhead{(mag)} & \colhead{(mJy)} & \colhead{} & \colhead{} & \colhead{(W\,Hz$^{-1}$)} & \colhead{} & \colhead{}}
\startdata
{TYC 7253-259-1} & 6180509689006780672 & Y & 74.8 & 1.22 & $0.28\pm0.14$ & 2.0 & 1.1:-1.8, 2.1:5.0, 3.1:0.7, 4.1:-1.5 & $8.2\times10^{8}$ & --- & --- \\
{TYC 3573-1391-1} & 2071061615589030400 & Y & 93.5 & 1.13 & $0.53\pm0.07$ & 7.8 & 1.1:3.3, 2.1:4.3, 3.1:2.9, 4.1:5.5 & $8.3\times10^{8}$ & --- & SE3.1 (1\farcs7), SE2.1 (1\farcs8), E1 (1\farcs8), QL4.1 (1\farcs8) \\
{1 Boo} & 1249218214732128640 & Y & 96.8 & 0.03 & $0.47\pm0.10$ & 4.7 & 1.2:2.4, 2.2:-0.1, 3.2:5.5 & $1.0\times10^{9}$ & SIMBAD & --- \\
{del Cap} & 6838311796136238976 & N & 11.6 & 0.69 & $0.58\pm0.10$ & 5.9 & 1.2:1.2, 2.2:7.3, 3.2:1.3 & $1.9\times10^{7}$ & SIMBAD; RUWE 3.9; $\Delta M_G=-1.37$ & --- \\
{HD 317101A} & 4050315620888089472 & N & 33.6 & 1.64 & $0.36\pm0.07$ & 4.9 & 1.1:-0.4, 2.1:0.0, 3.1:9.6, 4.1:1.1 & $2.0\times10^{8}$ & SIMBAD; Gaia NSS; RUWE 19.7 & QL3.1 (0\farcs4) \\
{109 Vir} & 3655377057091634304 & N & 41.2 & 0.06 & $0.48\pm0.09$ & 5.1 & 1.2:1.7, 2.2:3.8, 3.2:3.5 & $1.4\times10^{8}$ & RUWE 3.7; $\Delta M_G=-0.57$ & --- \\
{71 Tau} & 3312759303913978624 & N & 46.7 & 0.38 & $0.39\pm0.07$ & 5.6 & 1.1:1.8, 2.1:0.2, 3.1:2.7, 4.1:5.9 & $2.0\times10^{8}$ & SIMBAD; RUWE 5.0; $\Delta M_G=-1.31$ & --- \\
{V335 Peg} & 2717981134566374272 & N & 63.3 & 0.65 & $0.89\pm0.07$ & 12.5 & 1.1:1.2, 2.1:1.6, 3.1:3.8, 4.1:16.7 & $1.0\times10^{9}$ & SIMBAD; Gaia NSS; RUWE 5.9 & QL4.1 (0\farcs2) \\
{V2082 Cyg} & 2051422413763933568 & N & 87.6 & 0.50 & $0.18\pm0.07$ & 2.7 & 1.1:3.0, 2.1:-3.1, 3.1:5.1, 4.1:-0.2 & $5.8\times10^{8}$ & SIMBAD; $\Delta M_G=-0.95$ & --- \\
{12 Sco B} & 6042159790157494784 & N & 94.8 & 0.69 & $1.00\pm0.08$ & 13.0 & 1.1:3.9, 2.1:16.6, 3.1:3.3, 4.1:2.6 & $2.9\times10^{9}$ & SIMBAD; RUWE 4.1; $\Delta M_G=-0.92$ & --- \\
{HD 114993} & 6194009389692803584 & N & 95.1 & 0.36 & $0.51\pm0.09$ & 6.0 & 1.1:5.5, 2.1:1.4, 3.1:2.2, 4.1:2.7 & $9.4\times10^{8}$ & SIMBAD; RUWE 5.9; $\Delta M_G=-0.59$ & --- \\
\enddata
\tablecomments{$k$: Y = eligible star, counted in $k$; N = Gaia multiplicity evidence (RUWE\,$>1.4$, a non-single-star solution, or $\Delta M_G<-0.5$), so eligible only above $L_{\rm env}=10^{10}\,\WHz$, which none reaches (Section~\ref{sec:candidates}). $S_{\rm coadd}$, SNR: inverse-variance coadd over unvetoed epochs. In the steady test (coadd SNR\,$\ge5$) the counted forced detections give $k=\KSteady$ against $b=\BSteady$ expected at blank control positions under the same selection and veto; in the single-epoch test $k=\KSnap$ star-epochs against $b=\BSnap$. $L_{\nu,\max}$: spectral luminosity in the brightest epoch. Multiplicity: SIMBAD binary type, Gaia non-single-star solution (NSS), RUWE, and over-luminosity $\Delta M_G$ (mag) relative to the main-sequence ridge. Catalogs: independent per-epoch component catalogs with a $\ge5\sigma$ component within 4\arcsec\ of the star's position at that epoch, and the separation (E1: Epoch-1 QL; QL: NRAO QL lists; SE: Single Epoch lists; Med: QL median stack; Section~\ref{sec:catcheck}).}
\end{deluxetable*}
}

\subsection{Cross-check with per-epoch catalogs} \label{sec:catcheck}

We define secure associations from the Epoch-1 catalog alone, rather than from the union of the per-epoch catalogs, and rely on forced photometry for Epochs 2--4. The Epoch-1 cross-match only determines which detections are labeled secure and supplies the model for the confusion veto; the limits and the tests for unexplained emission use forced photometry, which measures every star in every epoch, including non-detections, with one estimator whose noise, false-positive rate, and recovery are calibrated at control positions. A union cross-match would therefore move stars between the secure associations and the forced detections without changing $\mu_{95}$ or the population limits. It would also combine heterogeneous catalogs (the published Epoch-1 catalog, interim lists for Epochs 3 and 4, and no complete list for Epoch~2), each requiring its own positional-error scale, flux bias, and chance-alignment calibration, and because a high-proper-motion star samples a different patch of background in each epoch, the number of chance trials per star would grow with the number of epochs. Epoch~1 has the poorest astrometry of the four epochs (Section~\ref{sec:vlass}), so a complete later epoch would arguably make a better primary catalog. To check that these choices do not change the results, we repeated the main steps with the \XcNCat\ independent component catalogs of Section~\ref{sec:vlass}. Each star-epoch was matched to the catalog of its own half-epoch, with the star propagated (proper motion and parallax) to the observation date of each component; a match is a $\ge5\sigma$ component within $2\farcs5$, and chance matches were counted in the same way at the eight control positions.

\emph{Flux scale and astrometry.} For compact matches with catalog S/N\,$>10$, our flux densities on the QL image scale agree with the QL catalog peaks and imply $\kappa=$\,\XcKappaRange\ in half-epochs 3.1--4.1, bracketing the adopted \KappaInj. The SE catalogs provide an independent flux scale. Our corrected flux densities are $\XcSEratio\pm\XcSEratioErr$ times the SE peaks ($N=\XcSEratioN$; \XcSEratioRange\ in individual half-epochs), and for the \XcSEoverQLN\ bright sources detected in both the QL and SE Epoch-3 catalogs the peaks agree, SE/QL\,$=\XcSEoverQL\pm\XcSEoverQLErr$. For these faint, compact sources the QL peaks therefore agree with the SE peaks, rather than being $\sim$8\% low as found for bright calibrators \citep{Kimball2026}; if the SE scale is correct, our later-epoch flux densities and limits are conservative by $\sim$\XcSEcons\% (Section~\ref{sec:caveats}). The scatter of catalog positions about the stars beyond the catalog errors is \XcAstSE (SE) and \XcAstQL (QL), comparable to the Epoch-1 positional errors after the astrometric correction (Section~\ref{sec:reliability}).

\emph{Sub-threshold flux scale.} Snapshot surveys can recover faint sources incompletely \citep{White2007}, and our limits rest on forced flux densities well below the catalog threshold. We therefore applied the same estimator, with the same calibration, to the \CosN\ of the \CosNcat\ VLA-COSMOS 3\,GHz sources \citep{Smolcic2017} that are unresolved at $0\farcs75$, single-component, and have no other catalog source within $8\arcsec$, in the QL images of half-epochs 1.1, 2.1, and 3.1 (\CosNmeas\ measurements). The inverse-variance-weighted ratio of our flux densities to the COSMOS values is $\CosRatioFaint\pm\CosRatioFaintErr$ for sources below 0.5\,mJy, $\CosLoRatio\pm\CosLoErr$ for \CosLoBin\,$\mu$Jy sources, whose individual S/N is $\sim$0.2, and \CosRatioMin--\CosRatioMax\ in bins from 0.05 to 1\,mJy. The estimator therefore shows no loss of faint flux and no dependence on flux density. The $\sim$10\% excess, at $2.4\sigma$, may come from a residual difference in flux scale or from fainter neighbors within the VLASS beam; we do not correct for it. If the excess were a bias in our measurements, our limits would be conservative by the same factor.

\emph{Detections.} Of the unvetoed star-epochs with forced S/N\,$\ge5$, \XcConcSEpct\% (\XcConcSE/\XcConcSEN) have an SE match and \XcConcQLpct\% (\XcConcQL/\XcConcQLN) a QL match; the misses lie mostly just above our threshold, where PyBDSF island detection and its local rms differ from our estimator. Conversely, \XcCatLtFive\ catalog matches have forced S/N\,$<5$; \XcCatLtFiveOff\ of them lie more than $1\farcs5$ from the star, and their number is consistent with the \XcCatChance\ chance matches expected from the control positions. \XcCandCat\ of the \NCandStars\ forced detections have a catalog counterpart in at least one epoch. The M dwarf G~133-48 is detected \asec{\GOTEsep} from the star in an independent catalog, and its cutout positions favor co-motion with the star (Section~\ref{sec:cand_results}). TYC~3573-1391-1, the one steady-test detection at an eligible star, is detected in four catalogs, always 1\farcs7--1\farcs8 west of the star, confirming that its emission is not coincident (Section~\ref{sec:cand_results}). Where the proper motion is large, the catalog positions follow the stars: \XcMoveAName\ and \XcMoveBName\ moved \asec{\XcMoveASpan} and \asec{\XcMoveBSpan} between the matched epochs, and a constant offset from the moving star fits far better than a fixed sky position ($\chi^2=\XcMoveACo$ and \XcMoveBCo\ against \XcMoveASt\ and \XcMoveBSt, for \XcMoveADof\ degrees of freedom each).

\emph{Confusion veto.} Replacing the Epoch-1 confusion veto by each epoch's own catalog (SE where available for Epoch~2, otherwise Epoch~1), and excluding components within $2\farcs5$ of the measurement position at both stars and controls, adds \XcOwnAdded\ forced detections (\XcOwnAddedUnexpl\ at eligible stars) whose Epoch-1 component was not a secure association or lay near the star's later position. It also raises the background, giving $k=\XcOwnKs$ against $b=\XcOwnBs$ ($\mu_{95}=\XcOwnMus$) in the steady test and $k=\XcOwnKe$ against $b=\XcOwnBe$ ($\mu_{95,{\rm snap}}=\XcOwnMue$) in the single-epoch test. Applying both vetoes gives $k=\XcUnionKs$ against $b=\XcUnionBs$ and $k=\XcUnionKe$ against $b=\XcUnionBe$, changing $\mu_{95}$ by $\XcMuSteadyChange\%$ and $\mu_{95,{\rm snap}}$ by $\XcMuSnapChange\%$. The headline limit, without background subtraction, depends on the veto only through $k$ and the number of stars: the own-epoch and combined vetoes give $f<\XcOwnFBz$ and $f<\XcUnionFBz$ at high luminosity ($\mu_{95}=\XcOwnMuBz$ and \XcUnionMuBz), against $\FBzElig$ for the adopted veto (below $L_{\rm env}$), so the least stringent variant weakens it by \XcOwnFBzPct\%. No variant has an excess; the largest, $k=\XcUnionKe$ against $b=\XcUnionBe$ in the single-epoch test with both vetoes, has $P(n\ge k\mid b)=\XcUnionPe$. The own-epoch variant has the highest background, $b=\XcOwnBs$ in the steady test, because its veto exempts every component within 2\farcs5 at both stars and controls, so that its $b$ includes every cataloged chance coincidence; its $k=\XcOwnKs$ is consistent with it ($P(n\le k\mid b)=\XcOwnPLow$; the eight control angles give \XcOwnAngleMin--\XcOwnAngleMax, and a bootstrap over stars gives $\pm\XcOwnBErrBoot$ for $b$). The background-free limits do not depend on $b$. The eligibility criteria matter more than the veto. Excluding AFGK stars from $\log L_X/L_{\rm bol}\ge-3.5$ instead of $-4$ (\NEligLxA\ eligible stars) gives $k=\KSteadyLxA$ against $b=\BSteadyLxA$ (steady) and $k=\KSnapLxA$ against $b=\BSnapLxA$ (single-epoch), and from $-4.5$ (\NEligLxB\ stars) $k=\KSteadyLxB$ against \BSteadyLxB\ and $k=\KSnapLxB$ against \BSnapLxB, so neither choice produces an excess; treating the AFGK multiples as eligible at all luminosities (\NEligWithMult\ stars) produces the excess described in Section~\ref{sec:cand_results}.

\emph{Stars found only in the catalogs, and the median stack.} \XcCatOnlyN\ stars that are neither explained secure associations nor forced detections have a $\ge5\sigma$ catalog component within $1\arcsec$, \XcCatOnlyMulti\ of them in two or more epochs. Three of these are X-ray-active (BD+40~883A, BD$-$07~2388, and V376~Cep, Section~\ref{sec:reliability}); the fourth, near 2MASS~J14223872+0646143, is the stationary source already identified as background by its NVSS and FIRST positions. Of the \NCatOnlySingle\ found in a single catalog, \NCatOnlySingleAct\ have activity evidence and \NCatOnlySingleElig\ are at eligible stars; the others are below our threshold in forced photometry or are confusion-vetoed, as expected for chance matches. BD+40~883A (15\,pc, $0\farcs65\,{\rm yr}^{-1}$) is detected in three epochs within $0\farcs6$ of its moving position, at $\sim$1\,mJy and 5--6$\sigma$, but at only 1.3--4.4$\sigma$ by forced photometry: our cutout noise estimate in this low-latitude Cygnus field ($b\approx-4^\circ$), $\sim$0.24\,mJy, exceeds the catalog rms of 0.16--0.18\,mJy, probably because of diffuse Galactic emission. In the QL median stack (rms $\sim\XcMedRms\,\mu$Jy) we tested the \XcMedN\ stars that move less than $1\arcsec$ over Epochs 1--3. \XcMedDet\ have a $\ge5\sigma$ component within $2\farcs5$. They comprise \XcMedSecure\ of the \XcMedSecureTested\ secure associations tested (most of the others are strongly variable, and the median suppresses flares), \XcMedExplained\ forced detections that are not counted and \XcMedUnexplained\ that are, \XcMedCatonly\ of the catalog-only stars, and \XcMedNone\ other stars. All \XcMedNone\ other matches lie \asec{\XcMedNoneSepMin}--\asec{\XcMedNoneSepMax} from the star, and they are fewer than the \XcMedChance\ chance matches expected. The deeper median stack therefore reveals no steady emitters missed by the coadd test.

\subsection{Limits on individual stars and on the population} \label{sec:limits}

\begin{figure}
\centering
\includegraphics[width=0.6\textwidth]{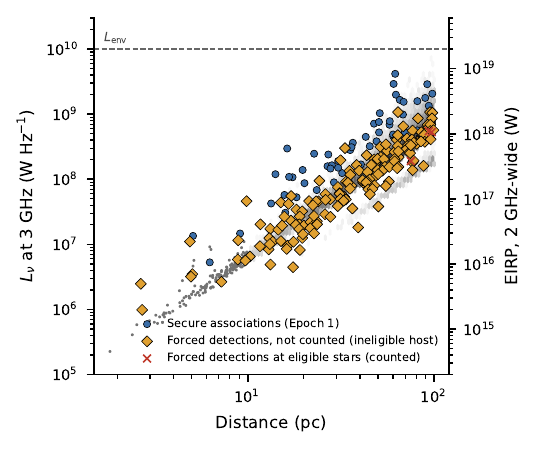}
\caption{All-epoch 95\% limits (completeness thresholds, or upper confidence bounds for stars with emission; Section~\ref{sec:limits_method}) on the 3\,GHz spectral luminosity at the positions of the sample stars (grey) versus distance. Colored symbols are measured luminosities, not limits: Epoch-1 catalog flux densities for the secure associations (blue), and all-epoch coadd flux densities for the forced detections that are not counted (orange: excluded stars, and M dwarfs and AFGK multiples below $L_{\rm env}$, dashed line) and counted (red: eligible stars); for the \NCandSubCoadd\ forced detections found only in single epochs the coadd is below $5\sigma$. The right axis gives the EIRP of a steady signal filling the 2\,GHz band.\label{fig:limits}}
\end{figure}

From the Epoch-1 catalog alone, the median 95\%-completeness limit is \SlimMed\,mJy. Combining all epochs, the median steady limit falls to \SlimAllMed\,mJy (5th--95th percentiles \SlimAllLo--\SlimAllHi\,mJy), corresponding to $\Lnu<\LlimAllMed\,\WHz$ (EIRP $<\EIRPlimAllMed$\,W across 2\,GHz) at the median distance, above the power of a Kardashev Type~I civilization \citep{Kardashev1964}, $\sim10^{16}$\,W in the planetary-scale definition of \citet{Gray2020}; the median single-snapshot limit is \SlimAnyMed\,mJy (\SlimBestMed\,mJy if the transmitter happened to be on in each star's lowest-noise epoch). Figure~\ref{fig:limits} shows the all-epoch limits against distance, and Table~\ref{tab:nearest} lists both sets for the twelve nearest stars. The most stringent, for Barnard's star, is $\Lnu<\BarnardLAll\,\WHz$ (EIRP $<\BarnardEIRPAll$\,W) from \BarnardNep\ epochs, compared with $\BarnardL\,\WHz$ from Epoch~1. These limits bound the total emission at each star's position, so at an undetected star they bound a transmitter whatever the star's eligibility, which governs only whether a detection could be recognized as one (Section~\ref{sec:candidates}).

\begin{figure}
\centering
\includegraphics[width=0.6\textwidth]{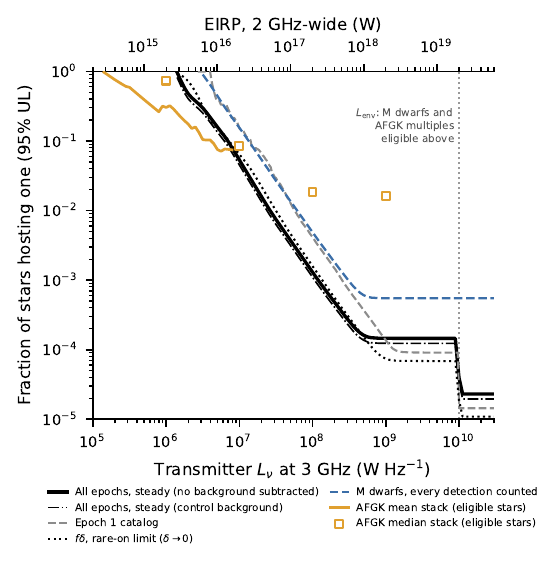}
\caption{95\% upper confidence limit on the fraction of eligible stars hosting a steady 2--4\,GHz transmitter above a given spectral luminosity (bottom axis) or 2\,GHz-wide EIRP (top axis): all epochs without background subtraction (solid black; the headline limit) and with the control-position background subtracted (dash-dotted), the Epoch-1 catalog analysis (dashed grey), and the rare-on limit on $f\delta$ ($\delta\to0$), counted per host star, without background subtraction (dotted; finite duty cycles are given in Section~\ref{sec:limits}). Well below $L_{\rm env}=10^{10}\,\WHz$ (vertical line) only the apparently single AFGK stars outside the excluded classes contribute; the M dwarfs and AFGK multiples enter over a narrow transition around $L_{\rm env}$, where a transmitter at one of them is detected with probability $\approx0.5$, so the limits drop. Blue dashed line: M dwarfs, with every coadd detection, natural emitters included, counted as a possible transmitter (no eligibility criteria). Orange line: approximate limit from the mean stacks of eligible AFGK stars for a population of transmitters with a common luminosity $L_T$, including host-sampling variance, shown where tighter than the individual limits; squares: median stacks of the pooled 0--100\,pc sample.\label{fig:pop}}
\end{figure}

Figure~\ref{fig:pop} shows the population limits. Counting the \KSteady\ steady-test detection at an eligible star as a possible transmitter ($\mu_{95}=\MuSteadyBz$, no background subtracted), fewer than $\FBzSeven$, $\FBzEight$, and $\FBzNine$ of the eligible AFGK stars host a steady 2--4\,GHz transmitter with $\Lnu$ above $10^7$, $10^8$, and $10^9\,\WHz$ respectively, and $f<\FBzElig$ up to $\sim0.8\,L_{\rm env}$, where the effective number of stars is the \NEligSteady\ eligible stars in the steady test. The M dwarfs and AFGK multiples then enter over a narrow transition: $f<\FBzAtEnv$ at $L_{\rm env}$ itself, where a transmitter at an envelope star is detected with probability $\approx0.5$, and $f<\FBzFloor$ above $\sim\FloorL\,\WHz$, where the effective number of stars is within 0.1\% of all \NAllFloor\ stars outside the excluded classes in the steady test; with no detection at all this would be $\FBzEnvKZero$. Without the multiplicity criterion, the \KSteadyWithMult\ steady-test detections among the \NWithMultSteady\ AFGK stars outside the excluded classes would give $\mu_{95}=\MuWithMult$ and $f<\FBzWithMult$ below $L_{\rm env}$, about twice the adopted value; because the criterion was adopted after these detections were examined (Section~\ref{sec:cand_results}), the adopted limit is conditional on it. Conversely, restricting the eligible stars to the \NXc\ in the steady test whose activity is measured below the X-ray threshold or lies below it at the nominal survey depth (Table~\ref{tab:xray}) removes TYC~3573-1391-1 and leaves $k=\KXc$, giving $f<\FXcNine$ above $10^9\,\WHz$ ($N_{\rm eff}=\NeffXcNine$) against $\FBzNineTwo$ for the adopted selection: the residual X-ray incompleteness barely changes the limit, because the smaller sample is offset by the smaller $k$. For the M dwarfs, counting every detection as a possible transmitter gives a limit that does not depend on any eligibility decision. Among the \NMallCoadd\ M dwarfs with a coadd, \KMall\ reach $5\sigma$ in the coadd (\KMallSec\ of them secure natural associations), so $\mu_{95}=\MuMall$ without background subtraction, and fewer than $\FMallSeven$, $\FMallEight$, and $\FMallNine$ of M dwarfs host a steady transmitter above $10^7$, $10^8$, and $10^9\,\WHz$ (effective numbers \NeffMallSeven, \NeffMallEight, and \NeffMallNine; Figure~\ref{fig:pop}). These limits reach three decades below $L_{\rm env}$, and $f<1$ at $\ReachMall\,\WHz$; the natural emitters counted in $k$ make them weaker than those for the eligible AFGK stars. Subtracting the control-position background ($\mu_{95}=\MuSteady$) would give $\FAllSeven$, $\FAllEight$, $\FAllNine$, $\FAllElig$ (well below $L_{\rm env}$), $\FAllAtEnv$ (at $L_{\rm env}$), and $\FAllFloor$. The Epoch-1 catalog analysis gives $f<\FpopFloor$ above $L_{\rm env}$, with $k=\KEOne$ unexplained secure associations and no background subtraction. For a transmitter on during a random fraction $\delta$ of snapshots, independently from snapshot to snapshot, counting host stars gives $f<\FDutyOneBz$, $\FDutyThreeBz$, and $\FDutyTenthBz$ above $\sim1.1\,L_{\rm env}$, beyond the transition, for $\delta=1$, 0.3, and 0.1 without background subtraction ($\FDutyOne$, $\FDutyThree$, and $\FDutyTenth$ after subtracting the control-position background), and $f\delta<\FDeltaFloorBz$ as $\delta\to0$ ($\FDeltaFloor$ background-subtracted).

\begin{deluxetable*}{lccrrrrrrrrr}
\tabletypesize{\footnotesize}
\tablecaption{95\% limits on steady 2--4\,GHz emission at the positions of the twelve nearest stars in the sample (completeness thresholds, or upper confidence bounds for stars with emission; Section~\ref{sec:limits_method}).\label{tab:nearest}}
\tablehead{\colhead{Star} & \colhead{Act.} & \colhead{Elig.} & \colhead{$d$} & \colhead{$G_{BP}-G_{RP}$} & \colhead{$S_{\rm lim}^{\rm E1}$} & \colhead{$L_{\nu,\rm lim}^{\rm E1}$} & \colhead{$N_{\rm ep}$} & \colhead{$S_{\rm coadd}$} & \colhead{$S_{\rm lim}$} & \colhead{$L_{\nu,\rm lim}$} & \colhead{EIRP$_{\rm lim}$} \\
\colhead{} & \colhead{} & \colhead{} & \colhead{(pc)} & \colhead{(mag)} & \colhead{(mJy)} & \colhead{(W\,Hz$^{-1}$)} & \colhead{} & \colhead{(mJy)} & \colhead{(mJy)} & \colhead{(W\,Hz$^{-1}$)} & \colhead{(W)}}
\startdata
{Barnard's star} & A & L & 1.83 & 2.83 & 1.27 & $5.1\times10^{5}$ & 3 & $+0.04\pm0.09$ & 0.57 & $2.3\times10^{5}$ & $4.5\times10^{14}$ \\
{Wolf 359} & A & L & 2.41 & 4.18 & 1.28 & $8.9\times10^{5}$ & 4 & $+0.18\pm0.09$ & 0.59 & $4.1\times10^{5}$ & $8.2\times10^{14}$ \\
{Lalande 21185} & A & L & 2.55 & 2.22 & 1.17 & $9.0\times10^{5}$ & 3 & $+0.02\pm0.11$ & 0.75 & $5.8\times10^{5}$ & $1.2\times10^{15}$ \\
{UV Cet} & A & L & 2.67 & 3.83 & 1.12 & $9.6\times10^{5}$ & 4 & $+2.90\pm0.08$ & 3.03 & $2.6\times10^{6}$ & $5.2\times10^{15}$ \\
{BL Cet} & A & L & 2.72 & 3.75 & 1.12 & $9.9\times10^{5}$ & 4 & $+1.11\pm0.08$ & 1.24 & $1.1\times10^{6}$ & $2.2\times10^{15}$ \\
{Ross 154} & A & L & 2.98 & 2.83 & 1.20 & $1.3\times10^{6}$ & 3 & $+0.14\pm0.10$ & 0.63 & $6.7\times10^{5}$ & $1.3\times10^{15}$ \\
{Ross 248} & A & L & 3.16 & 3.53 & 1.01 & $1.2\times10^{6}$ & 3 & $-0.20\pm0.08$ & 0.53 & $6.4\times10^{5}$ & $1.3\times10^{15}$ \\
{eps Eri} & A & L & 3.22 & 1.14 & 1.11 & $1.4\times10^{6}$ & 3 & $-0.11\pm0.10$ & 0.66 & $8.2\times10^{5}$ & $1.6\times10^{15}$ \\
{Lacaille 9352} & A & L & 3.29 & 2.10 & 0.95 & $1.2\times10^{6}$ & 4 & $-0.06\pm0.07$ & 0.47 & $6.1\times10^{5}$ & $1.2\times10^{15}$ \\
{Ross 128} & A & L & 3.37 & 3.03 & 1.39 & $1.9\times10^{6}$ & 4 & $+0.09\pm0.08$ & 0.56 & $7.7\times10^{5}$ & $1.5\times10^{15}$ \\
{61 Cyg B} & A & E & 3.50 & 1.72 & 0.87 & $1.3\times10^{6}$ & 4 & $+0.03\pm0.07$ & 0.43 & $6.3\times10^{5}$ & $1.3\times10^{15}$ \\
{61 Cyg A} & A & E & 3.50 & 1.46 & 0.87 & $1.3\times10^{6}$ & 4 & $+0.03\pm0.07$ & 0.43 & $6.3\times10^{5}$ & $1.3\times10^{15}$ \\
\enddata
\tablecomments{Act.: A = activity evidence (an X-ray counterpart or a SIMBAD activity type; Section~\ref{sec:candidates}), which at these distances is found for almost every star and is not used to select technosignature hosts. Elig.: technosignature eligibility (Section~\ref{sec:candidates}); E = eligible at any luminosity, L = eligible only above $L_{\rm env}=10^{10}\,\WHz$ (M dwarfs, and AFGK stars with Gaia multiplicity evidence), X = excluded (known radio-emitter class or $\log L_X/L_{\rm bol}\ge-4$). Every limit bounds the total emission at the star's position, natural or artificial, whatever its eligibility. E1: Epoch-1 catalog analysis. $N_{\rm ep}$, $S_{\rm coadd}$ (measured coadd flux density and its error), $S_{\rm lim}$, $L_{\nu,\rm lim}$: all-epoch coadded forced photometry. EIRP$_{\rm lim}=L_{\nu,\rm lim}\times2$\,GHz (steady, band-filling signal). The limits apply to emission within about one beam radius of the star; 1\arcsec\ corresponds to $d$\,au projected, and a transmitter on a planet at a wider separation would lie outside the measured position.}
\end{deluxetable*}

\subsection{Stacked emission} \label{sec:stacks}

\begin{figure*}
\centering
\includegraphics[width=0.95\textwidth]{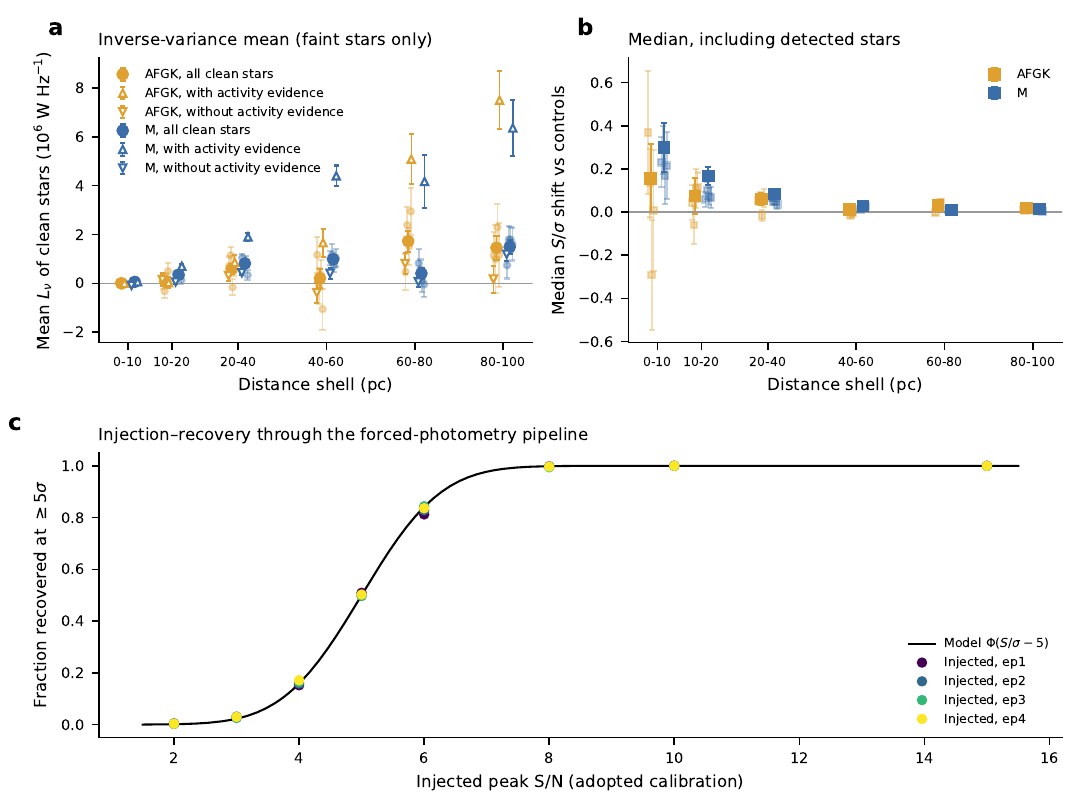}
\caption{(a) Inverse-variance-weighted mean 3\,GHz spectral luminosity of clean stars in distance shells, for the all-epoch coadd (large circles) and individual epochs (small circles), and for the all-epoch coadd split into stars with (open upward triangles) and without (open downward triangles) activity evidence. (b) Shift of the median $S/\sigma$ relative to control positions, including detected stars. (c) Fraction of injected point sources recovered at $\ge5\sigma$ against their true $S/\sigma$, compared with $\Phi(S/\sigma-5)$.\label{fig:stacks_all}}
\end{figure*}

Table~\ref{tab:stacks_all} and Figure~\ref{fig:stacks_all} present the all-epoch stacks. The M-dwarf mean exceeds $4\sigma$ in \NMShellsFour\ of the six shells and is most significant at 20--40\,pc, $\langle\Lnu\rangle=(\AllMSL\pm\AllMSLErr)\,\WHz$ (\AllMSNR$\sigma$, \AllMN\ stars, $N_{\rm eff,w}=\AllMNeffW$), and the Epoch-1 image stack of the same shell gives $\langle\Lnu\rangle=(\StackMSL\pm\StackMSLerr)\,\WHz$ (\StackMSLsnr$\sigma$), higher because it excludes only the cataloged detections and so retains the forced detections, including Epoch-1 flares, that the all-epoch stack excludes. Splitting the clean stars by activity evidence (Table~\ref{tab:stacks_all}) shows that most of this signal comes from the minority with it: at 20--40\,pc the \AllMNAct\ M dwarfs with activity evidence (\ActFracMForty\% of the clean stars) give $\langle\Lnu\rangle=(\AllMSLAct\pm\AllMSLErrAct)\,\WHz$ (\AllMSNRAct$\sigma$), against $(\AllMSLIn\pm\AllMSLErrIn)\,\WHz$ (\AllMSNRIn$\sigma$) for the \AllMNIn\ without. The M dwarfs without activity evidence reach at most \InMmaxSNR$\sigma$ in any shell; their weaker signal may come from stars whose activity is not cataloged. The median stack, which includes the detected stars, shows the same excess, $\Delta\,{\rm med}(S/\sigma)=\AllMDmed\pm\AllMDmedErr$ (\AllMDmedSNR$\sigma$), but a median shift cannot distinguish faint emission from every star from brighter emission from a minority: a shift of \AllMDmed\ is produced equally by every star at that fraction of $\sigma$ or by 6--10\% of stars at or well above $1\sigma$. The stacked signal is therefore population-averaged emission concentrated in the stars with activity evidence, as expected if radio emission traces magnetic activity \citep{GudelBenz1993,Gudel2002,Callingham2021,Yiu2024}, rather than a level common to all M dwarfs. The AFGK means reach at most \AllAFGKmaxSNR$\sigma$ (\AllAFGKmaxShell\,pc), \ActAFGKmaxSNR$\sigma$ for the stars with activity evidence, and \InAFGKmaxSNR$\sigma$ for those without. In every shell with more than 1000 clean stars, the ten highest-weight stars carry at most \TopTenMax\% of the weight, so no stack is dominated by a few nearby stars. The rise of $\langle\Lnu\rangle$ with distance reflects the exclusion of detected stars, which truncates each shell at a luminosity proportional to $d^2$. Figure~\ref{fig:stacks_all}c shows the injection--recovery test of Section~\ref{sec:forced}.

\begin{deluxetable*}{llrrrrrrrrr}
\tabletypesize{\footnotesize}
\tablecaption{Stacks of the all-epoch coadded forced photometry.\label{tab:stacks_all}}
\tablehead{\colhead{Shell} & \colhead{Class} & \colhead{$N_{\rm all}$} & \colhead{$N_{\rm clean}$} & \colhead{$\langle L_\nu\rangle$} & \colhead{SNR} & \colhead{$\langle L_\nu\rangle_{\rm act}$} & \colhead{$\langle L_\nu\rangle_{\rm no\,act}$} & \colhead{$N_{\rm eff,w}$} & \colhead{Top 10} & \colhead{$\Delta\,{\rm med}(S/\sigma)$} \\
\colhead{(pc)} & \colhead{} & \colhead{} & \colhead{} & \colhead{($10^5$ W\,Hz$^{-1}$)} & \colhead{} & \colhead{($10^5$ W\,Hz$^{-1}$)} & \colhead{($10^5$ W\,Hz$^{-1}$)} & \colhead{} & \colhead{(\%)} & \colhead{}}
\startdata
0--10 & AFGK & 44 & 44 & $0.0\pm0.4$ & $0.1$ & $0.0\pm0.4$ & \nodata & 9 & 76.2 & $+0.157\pm0.156$ \\
0--10 & M & 165 & 152 & $0.6\pm0.2$ & $2.7$ & $0.7\pm0.2$ & $-0.8\pm0.8$ & 9 & 65.4 & $+0.299\pm0.114$ \\
10--20 & AFGK & 318 & 312 & $0.6\pm1.6$ & $0.4$ & $0.3\pm2.1$ & $1.5\pm2.5$ & 157 & 15.3 & $+0.075\pm0.082$ \\
10--20 & M & 1{,}229 & 1{,}194 & $3.5\pm0.7$ & $4.8$ & $7.0\pm0.9$ & $0.6\pm0.9$ & 599 & 4.6 & $+0.168\pm0.042$ \\
20--40 & AFGK & 2{,}713 & 2{,}694 & $5.7\pm2.0$ & $2.8$ & $8.6\pm3.1$ & $3.2\pm2.2$ & 1{,}134 & 3.8 & $+0.060\pm0.027$ \\
20--40 & M & 9{,}897 & 9{,}861 & $8.1\pm0.9$ & $8.8$ & $19.0\pm1.7$ & $4.3\pm1.1$ & 3{,}652 & 2.7 & $+0.084\pm0.014$ \\
40--60 & AFGK & 7{,}732 & 7{,}697 & $2.0\pm3.8$ & $0.5$ & $16.6\pm5.7$ & $-3.8\pm4.1$ & 4{,}011 & 2.3 & $+0.013\pm0.016$ \\
40--60 & M & 25{,}641 & 25{,}627 & $9.9\pm1.8$ & $5.4$ & $44.0\pm4.3$ & $4.1\pm2.2$ & 14{,}156 & 0.9 & $+0.028\pm0.008$ \\
60--80 & AFGK & 14{,}234 & 14{,}206 & $17.3\pm4.3$ & $4.0$ & $50.8\pm10.1$ & $8.1\pm4.3$ & 7{,}971 & 1.2 & $+0.030\pm0.012$ \\
60--80 & M & 47{,}527 & 47{,}512 & $4.1\pm2.1$ & $1.9$ & $41.7\pm10.7$ & $0.6\pm2.2$ & 29{,}033 & 0.4 & $+0.009\pm0.006$ \\
80--100 & AFGK & 23{,}232 & 23{,}210 & $14.6\pm5.0$ & $2.9$ & $75.0\pm12.0$ & $1.7\pm5.5$ & 14{,}584 & 0.7 & $+0.019\pm0.008$ \\
80--100 & M & 75{,}482 & 75{,}464 & $14.9\pm2.8$ & $5.4$ & $63.6\pm11.5$ & $11.9\pm2.9$ & 46{,}937 & 0.2 & $+0.015\pm0.005$ \\
\enddata
\tablecomments{$\langle L_\nu\rangle$: inverse-variance-weighted mean spectral luminosity of clean stars (no secure association, no forced detection, no epoch with predicted catalog contamination $\ge0.5\sigma$), after subtracting the mean at the control positions; the error is the larger of the formal error and the scatter among the eight control positions. $N_{\rm eff,w}=(\sum w)^2/\sum w^2$; Top 10: weight fraction carried by the ten highest-weight stars. $\langle L_\nu\rangle_{\rm act}$ and $\langle L_\nu\rangle_{\rm no\,act}$: the same mean for the clean stars with and without activity evidence (Section~\ref{sec:candidates}); \nodata: fewer than ten stars. The control positions are subject to the same clean cuts as the stars. $\Delta\,{\rm med}(S/\sigma)$: median $S/\sigma$ of all unvetoed stars in the shell, including detected stars, minus the median at control positions; bootstrap error.}
\end{deluxetable*}

Treating the entire stacked signal of the eligible AFGK stars as potentially artificial, and assuming that every host has the same luminosity $L_T$ (a brighter host would be individually detected and removed from the clean stack, so these are approximate limits for a population of transmitters with a common luminosity), the mean stacks reach $f<1$ at $L_T=\ReachAllAFGK\,\WHz$, \ReachRatioAFGK\ times fainter than the luminosity at which the individual detections reach $f<1$ ($\ReachIndBz\,\WHz$), and give $f<\FMeanSixAFGK$ at $10^6\,\WHz$ (Figure~\ref{fig:pop}). These limits describe few stars: at $10^6\,\WHz$ the effective number of stars is \NeffSixAFGK, and stars within 10\,pc carry \WtenSixAFGK\% of the weight, so they constrain the nearest stars rather than the population; the host-sampling variance is included. Injecting transmitters at $L_T=1.5\times10^5$--$10^7\,\WHz$, just above the $f<1$ crossing, with $f=0.1$--1 into the measurements at the control positions, removing each pseudo-star whose coadd or any single epoch then reaches $5\sigma$, as for the real stars, and applying the same selection and estimator, the limit covers the injected $f$ in at least \CovAFGKmin\ of the \NCovTrials\ trials in each case. These trials combine 25 random host assignments with each of only \CovNangles\ noise realizations, one per control angle, and the shortfalls come mainly from a single realization, whose mean lies within $1.3\sigma$ of the other seven. Eight realizations cannot establish 95\% coverage, so we regard the stack limits, particularly near $f=1$, as approximate. The median stack of the pooled eligible stars bounds the fraction directly, independent of whether the transmitters are individually detected: $f<\FMedSevenAFGK$ at $10^7\,\WHz$ and $f<\FMedNineAFGK$ at $10^9\,\WHz$. The individual limits are more constraining than the mean stacks above $\approx\StackCrossAFGK\,\WHz$, and than the median stack at $10^7\,\WHz$ and above. We derive no transmitter limits from the M-dwarf stacks: at these luminosities, far below $L_{\rm env}$, a transmitter at an M dwarf could not be distinguished from the natural emission that the stacks detect.

\section{Discussion} \label{sec:discussion}

\subsection{Comparison with narrowband searches} \label{sec:bl}

\begin{figure*}
\centering
\includegraphics[width=\textwidth]{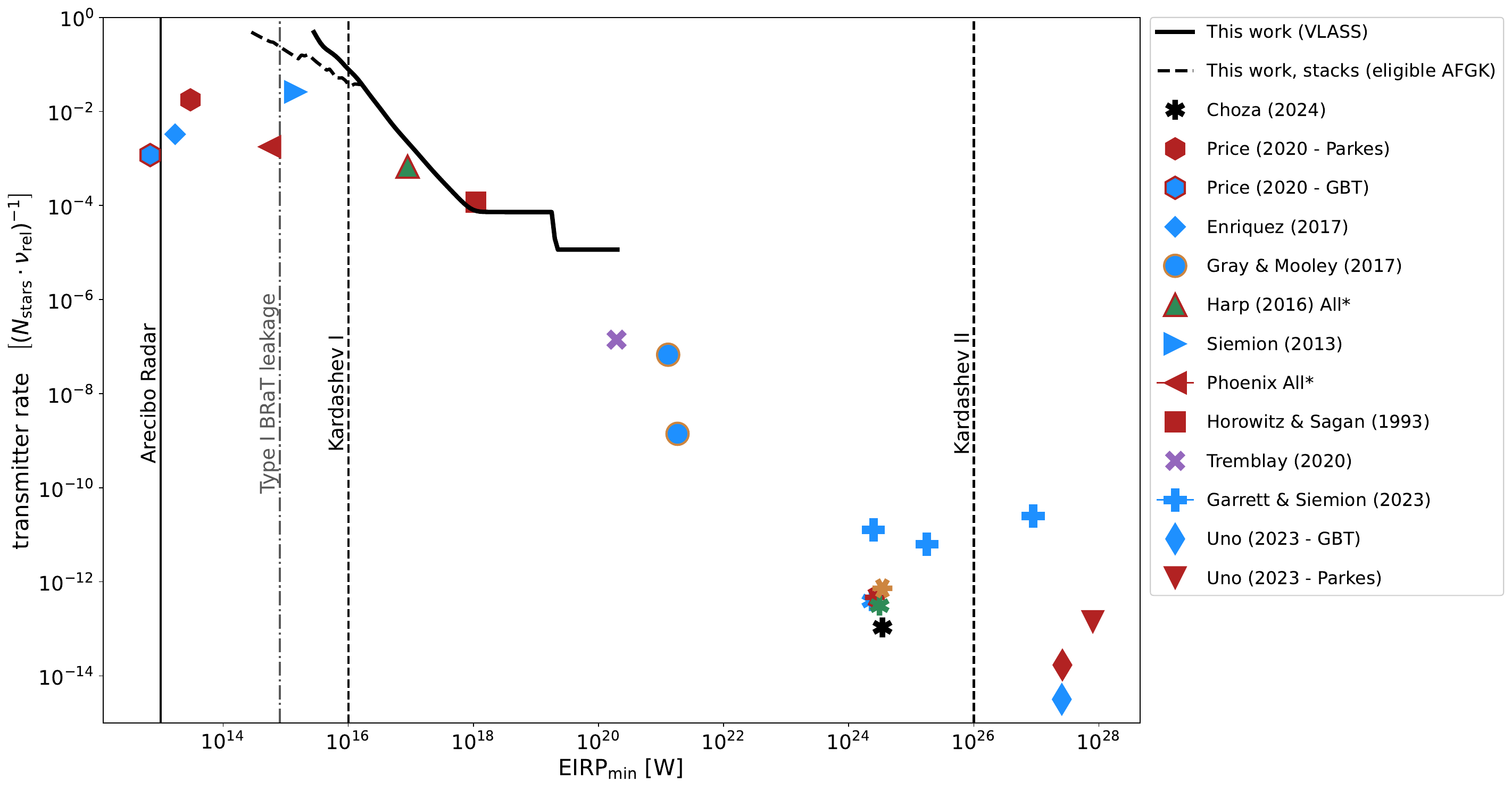}
\caption{Transmitter rate versus EIRP$_{\rm min}$ for this work and for previous radio technosignature surveys \citep{HorowitzSagan1993,Backus2004,Siemion2013,Harp2016,Enriquez2017,GrayMooley2017,Price2020,TremblayTingay2020,Garrett2022,Uno2023,Choza2024}; literature values are plotted with the compilation of B.~Jacobson-Bell et al.\ (in preparation). Narrowband EIRP$_{\rm min}$ values are the total power of a transmitter narrower than each search's frequency channel (0.05--3\,Hz for most searches, and 15 and 122\,Hz for \citealt{GrayMooley2017}); for the MWA survey we plot the 10\,Hz transmission-bandwidth case of \citet{TremblayTingay2020}, their limits for 10\,kHz channels raised by their factor of 1000. The EIRP$_{\rm min}$ values of the searches that used turboSETI \citep{Enriquez2017,Price2020,Garrett2022,Uno2023} are increased by the factor of 3.3 found by \citet{Choza2024}, whose own limits include it; the other searches did not use turboSETI and are plotted as published. VLASS limits (black) are for the total in-band power of a steady signal spread over many channels: individual limits without background subtraction (solid; the step at $2\times10^{19}$\,W is where the M dwarfs and AFGK multiples become eligible) and, where more constraining, the approximate mean-stack limit for eligible AFGK stars (dashed), which assumes a common transmitter luminosity. Vertical lines mark the EIRP of the Arecibo radar, the Kardashev Type~I and II power levels \citep{Kardashev1964}, and the Type~I broadband leakage benchmark of \citet{Garrett2026} spread over 2\,GHz.\label{fig:tr}}
\end{figure*}

Table~\ref{tab:bl} compares our limits for a star at 50\,pc with those of \citet{Price2020}, whose GBT and Parkes bands overlap VLASS at 2.0--2.8 and 2.6--3.45\,GHz. In the text and Table~\ref{tab:bl} the narrowband limits include the factor of 3.3 correction of \citet{Choza2024} to the turboSETI signal-to-noise ratios, and Table~\ref{tab:bl} also lists the published values; Figure~\ref{fig:tr} applies the same correction to all the searches that used turboSETI. In flux density VLASS is $\sim\FluxRatio$ times deeper (\VlassFiftyS\,mJy against $\sim\BLSJy$\,Jy), because it integrates over 2\,GHz rather than a 2.79\,Hz channel. For a narrow line, however, the continuum dilutes the power over the band, and our in-band EIRP limit ($\VlassFiftyEIRP$\,W) is $\sim\BLratio$ times higher than the BL limit of $\BLEIRPcorr$\,W ($\sim\BLratioPub$ times the published $\BLEIRPpub$\,W). A search matched to a signal of bandwidth $B$ in the same narrowband data would lose sensitivity to total power only as $\sqrt{B}$, so for any bandwidth within the band a dedicated search of the GBT data would remain the more sensitive. The advantage of the continuum survey is coverage rather than depth: the turboSETI pipeline \citep{Enriquez2019} used by \citet{Price2020} is built for drifting tones and does not search for radar chirps, spread-spectrum links, or broadband leakage (although \citealt{Gajjar2022} searched the BL data for dispersed broadband pulses), whereas the continuum measurement includes them, provided that they survive the interference flagging (Section~\ref{sec:limits_method}), for every star in the sample. In occurrence, and in a common construction (a 50\% detection probability per star, as adopted by \citealt{Price2020}, and no background subtraction; Table~\ref{tab:bl}), the \NFifty\ eligible stars within 50\,pc give $f<\FFiftyHalf$, about \FFiftyRatio\ times tighter than the \BLfCommon\% for the \NGBTS\ GBT S-band targets of \citet{Price2020}, who quote 0.37\%, albeit at an EIRP threshold $\sim\BLratio$ times higher. The two approaches are complementary: targeted narrowband searches of nearby stars constrain faint, narrow, long-duty-cycle signals around $\sim10^3$ stars, and a continuum survey constrains bright, broad signals of any waveform around every one of the $\sim10^5$ stars in the volume. Narrowband searches of whole survey fields \citep[e.g.,][]{Wlodarczyk2020}, and the commensal searches now running at MeerKAT and the VLA \citep{Czech2021,Tremblay2025}, reach larger stellar samples, on average of more distant stars. Whatever their overlap with our volume, however, a search for narrowband signals alone would in most cases miss a BRaT, so narrowband and continuum searches remain complementary.

\begin{deluxetable*}{llllrrr}
\tablecaption{Comparison with the Breakthrough Listen narrowband survey of \citet{Price2020} for a star at 50\,pc.\label{tab:bl}}
\tablehead{\colhead{Survey} & \colhead{Band} & \colhead{Dwell} & \colhead{$S_{\rm lim}$ (Jy)} & \colhead{EIRP$_{\rm lim}$ (W)} & \colhead{$N_\star$} & \colhead{$f_{95}$ (\%)}}
\startdata
BL GBT & 1.80--2.80 GHz & $3\times5$ min ON & $2.3\times10^{1}$ & $6.9\times10^{12}$ ($2.1\times10^{12}$) & 1{,}005 & $0.6$ \\
BL Parkes & 2.60--3.45 GHz & multi-epoch & $1.0\times10^{2}$ & $3.0\times10^{13}$ ($9.1\times10^{12}$) & 189 & $3.2$ \\
This work (VLASS QL) & 2--4 GHz & $\sim$5 s snapshots & $5.4\times10^{-4}$ & $3.2\times10^{17}$ & 4{,}103 & $0.15$ \\
\enddata
\tablecomments{Breakthrough Listen (BL) limits apply to signals $\lesssim$3\,Hz wide with $|\dot\nu|\le4$\,Hz\,s$^{-1}$ and use the 1\,Hz transmitter-width convention of \citet{Price2020}, including the factor of 3.3 correction of \citet{Choza2024} (published values in parentheses), as in Figure~\ref{fig:tr}; the VLASS limit applies to the total in-band power of signals that survive the interference flagging (Section~\ref{sec:limits_method}). $N_\star$ is the number of stars constraining the occurrence limit (for BL GBT, the S-band targets, the band that overlaps VLASS; for VLASS, stars within 50\,pc without activity evidence). $f_{95}$ is the 95\% upper limit on the fraction of stars hosting a transmitter above EIRP$_{\rm lim}$ in a common construction: a 50\% detection probability per star, as adopted by \citet{Price2020}, and no background subtraction ($3/0.5N_\star$ for no detections; for VLASS $\mu_{95}=\MuFifty$ for the \KFifty\ steady-test detections among these stars). \citet{Price2020} quote 0.37\% (GBT S band) and 2.0\% (Parkes).}
\end{deluxetable*}

Figure~\ref{fig:tr} places these limits in the plane of transmitter rate against EIRP$_{\rm min}$ used to compare technosignature surveys \citep{Enriquez2017,Price2020,Painter2025}. The transmitter rate is $(N_\star\,\nu_{\rm rel})^{-1}$, where $\nu_{\rm rel}=\Delta\nu/\nu_c$ is the fractional bandwidth, 2/3 for VLASS; with $k>0$ we use the equivalent $f_{95}/(3\nu_{\rm rel})$, which reduces to it for $k=0$. Because our population limit at a given EIRP involves only the $N_{\rm eff}$ stars individually sensitive to it, the VLASS limit is a curve rather than a point. It falls from 0.5 at $\ReachIndEIRP$\,W, where $f=1$, to $\TRfloor$ above $\sim\FloorEIRP$\,W, where the effective number of stars is within 0.1\% of all \NAllFloor\ stars outside the excluded classes; below $L_{\rm env}$ (EIRP $2\times10^{19}$\,W) the limit rests almost entirely on the eligible AFGK stars, and the envelope stars enter over a narrow transition around it. For the stacks, which limit a fraction rather than count stars, we use the equivalent number $N=3/f_{95}$ and show them only where they are more constraining than the individual limits. The VLASS floor is lower than the transmitter rate reached by any of the surveys shown with EIRP$_{\rm min}<10^{20}$\,W, but the comparison is not like for like. The narrowband EIRP$_{\rm min}$ values are for transmitters narrower than each search's channel, typically a few hertz, whereas the VLASS values are the total in-band power of a steady signal spread over many channels that survives flagging and imaging (Section~\ref{sec:limits_method}).

\subsection{Comparison with broadband technosignature benchmarks} \label{sec:brats}

Treating the band-averaged 3\,GHz luminosity as representative of a flat broadband spectrum, the Type~I leakage benchmark of \citet{Garrett2026}, $\Lnu\sim4\times10^5\,\WHz$ (EIRP $\sim8\times10^{14}$\,W across 2\,GHz), is reached individually only for Barnard's star, whose all-epoch limit is $\Lnu<\BarnardLAll\,\WHz$. Barnard's star is undetected, so this bounds the total emission at its position and excludes a Type~I BRaT there. A detection at that level, however, could not have been attributed to a transmitter, because Barnard's star is an M dwarf and natural emission could produce it. The same holds for the M dwarfs that make up most of the nearest stars: continuum non-detections bound BRaTs at each of them, but a continuum detection alone cannot reveal one, and the further diagnostics proposed by \citet{Garrett2026} would be required. The stacks of eligible AFGK stars just reach the benchmark: for transmitters that all have $\Lnu=4\times10^5\,\WHz$ they require $f<\FBratAllAFGK$, but with \NeffBratAFGK\ effective stars, of which those within 10\,pc carry \WtenBratAFGK\% of the weight, this constrains the nearest AFGK stars rather than the population. Earth's present aggregate leakage of $\sim400\,\WHz$ would be detectable at the median all-epoch sensitivity only to $\sim\EarthLeakAUAll$\,au \citep[see also][]{Sheikh2025}. Several of the diagnostics proposed by \citet{Garrett2026} are out of reach at VLASS resolution and depth---brightness temperature, polarization, spectral structure, and scintillation---but two are central to this work: association with a Gaia star, and co-motion with it (Sections~\ref{sec:lspm} and \ref{sec:multiepoch}). A BRaT from a planet would be offset from its host by $\lesssim0\farcs1$ for orbits of a few au at $d\gtrsim30$\,pc, well within our matching tolerance. Frequencies $\gtrsim1$\,GHz also avoid the elevated ionospheric cut-off expected for planets around active M dwarfs \citep{Garrett2026,GarrettIono2026}, which favors 2--4\,GHz for the M dwarfs that make up most of our sample.

\citet{Zuckerman2026} proposed a benchmark of a different kind: a 60\,MW transmitter 200\,pc away that beams continuously toward Earth, delivering $\sim10^{10}$\,Jy in a channel a few Hz wide. Averaged over our 2\,GHz band this is $\sim\ZuckSband$\,Jy, or $\Lnu\sim\ZuckL\,\WHz$ (EIRP toward Earth $\sim\ZuckEIRP$\,W), more than $\ZuckMarginMed$ times our median limit and more than $\ZuckMarginWorst$ times the least sensitive per-star limit ($\Lnu<\LlimAllMax\,\WHz$). Because our limits constrain the flux density received at Earth, they apply directly to beamed transmission; only their translation into transmitter power assumes isotropy. Every star with forced photometry therefore excludes this power individually, for a transmitter operating within 2--4\,GHz while VLASS observed and spreading it over many channels; the luminosity is far above $L_{\rm env}$, so every star outside the excluded classes is eligible, M dwarfs included, and since the signal is continuous the steady test applies, giving $f<\FBzFloor$ at this luminosity. The benchmark itself is narrowband, however, and \citet{Zuckerman2026} placed it at 10\,GHz, so ours is an in-band analogue. Concentrated in one 2\,MHz VLASS channel the same signal would reach $\sim\ZuckChan$\,Jy, an extreme outlier that the automated flagging would almost certainly remove (Section~\ref{sec:limits_method}); excluding the narrowband benchmark itself requires imaging the native channels with flagging that preserves narrow, persistent signals (Section~\ref{sec:prospects}). Zuckerman's target class, old F6--M2 dwarfs, corresponds to $\BPRPlo\le G_{BP}-G_{RP}\le\BPRPhi$ \citep{Pecaut2013}. Of the \NSolarSingle\ apparently single stars in this range with all-epoch limits (RUWE$\,<1.4$, no Gaia non-single-star solution, not over-luminous), \NSolarActive\ are excluded or not in the steady test, and \NSolarUnexpl\ are counted forced detections (TYC~3573-1391-1, whose emission is offset from the star, and TYC~7253-259-1, detected in a single epoch only). Only TYC~3573-1391-1 passes the steady test, which applies to a continuous signal, so even if it were a transmitter $f<\FSolarSingle$ for the other \NSolarInactive\ ($k=\NSolarSteady$, without background subtraction). We have no ages for these stars. If a fraction $q$ of them are old, the limit for the old subset alone is weaker, $f_{\rm old}<f/q$; at the benchmark luminosity itself, however, age does not matter, because every star is excluded individually. Zuckerman's argument that any radio detection of an old solar-type star would be notable \citep[citing][]{Matthews2025} holds only after activity is excluded: active solar-type stars are detected in VLASS \citep[][and this work]{Davis2026}, which is why we exclude the known emitter classes and saturated coronae and treat multiples with caution (Section~\ref{sec:candidates}).

\subsection{Interpretation and caveats} \label{sec:caveats}

Our limits apply to emission that is on during the few-second VLASS snapshots (up to \NEpochTags\ per star between 2017 and 2026), in Stokes~I, within the 1.965--4.013\,GHz correlator band, excluding data flagged for radio-frequency interference (the spectral windows at 2.093--2.221\,GHz and 3.629--4.013\,GHz, which contain satellite downlinks, are the most heavily affected, so a transmitter confined to them is less well constrained; \citealt{Kimball2026}), and within about one beam radius of the star (Table~\ref{tab:nearest}). The EIRP values assume a flat, band-filling spectrum; for a signal occupying a fraction of the band the EIRP limit is unchanged if the signal survives the flagging and calibration weights at the frequencies it occupies (Section~\ref{sec:limits_method}), whereas the spectral-luminosity limit applies to the band-averaged power. In the Epoch-1 catalog analysis our noise estimate (the mean island rms of each image) is slightly conservative; the all-epoch limits use the calibrated cutout rms (Section~\ref{sec:forced}).

The sample is defined by Gaia DR3 astrometry and photometry ($\varpi/\sigma_\varpi>10$, a measured $G_{BP}-G_{RP}$, and the main-sequence window). It contains \PctGCNS\% of the GCNS stars within 100\,pc in the VLASS sky (Section~\ref{sec:gaia}), omits the faintest late-M and substellar objects, and includes unresolved binaries. It also omits \NBrightMissing\ bright ($V<4$) dwarfs within 100\,pc and north of $\delta=-40^\circ$, among them Sirius, Vega, Altair, Fomalhaut, and Castor, which lack a Gaia DR3 source or a precise parallax, or which fall outside the main-sequence window. We measured these at their SIMBAD positions with the same forced photometry: \NBrightUndet\ are undetected (all below $\BrightUndetMaxSNR\sigma$), with limits of $\BrightLSirius$, $\BrightLAltair$, $\BrightLVega$, and $\BrightLFomalhaut\,\WHz$ for Sirius, Altair, Vega, and Fomalhaut, and the \NBrightDet\ detections are Algol ($\beta$~Per, \AlgolS\,mJy), the magnetic Ap star $\theta$~Aur (\ThetAurSNR$\sigma$), and the multiple system $\xi$~Tau (\XiTauSNR$\sigma$), all plausible natural emitters. They are not included in the population limits, which apply to the Gaia sample, with each star weighted by its own sensitivity, rather than to a complete volume.

The classes in Table~\ref{tab:matches} follow SIMBAD; some stars classed as X-ray-active have more specific types in the literature ($\sigma$~CrB~A is an RS~CVn binary, and RZ~Cas an Algol-type eclipsing binary), which does not affect the analysis. The classification of natural emitters relies in part on SIMBAD, a live database, and the flux scale carries a \KappaSys\% systematic uncertainty from the peak-loss calibration. Our flux scale follows the calibrator-based QL biases of \citet{Kimball2026}; the Single Epoch catalogs suggest that the later-epoch QL peaks of faint compact sources are less biased, in which case our later-epoch limits are conservative by up to $\sim$\XcSEcons\% (Section~\ref{sec:catcheck}). Injection into the restored images (Section~\ref{sec:forced}) tests the estimator but not the imaging, which in snapshot surveys can recover faint flux incompletely \citep{White2007}; the VLA-COSMOS comparison of Section~\ref{sec:catcheck} shows that our estimator recovers sources down to $\sim$20\,$\mu$Jy with the calibrated flux scale.

Finally, a transmitter can be recognized only where it would stand out from natural emission, and the eligibility criteria of Section~\ref{sec:candidates} set where that is. Well below $L_{\rm env}$ the population limits rest on the \NElig\ eligible AFGK stars, \PctElig\% of the sample, because an M-dwarf or AFGK-multiple detection counts only above $L_{\rm env}$, and the \NExclAll\ stars of the known-emitter classes or with saturated coronae never count. The criteria are intrinsic properties of the stars, unlike the presence of activity evidence, which would have removed \PctActTen\% of the stars within 10\,pc but only \PctActHundred\% at 80--100\,pc, and the net eligible fraction of AFGK stars is nearly flat with distance (Section~\ref{sec:candidates}). Whether a star can be shown to lie below the X-ray threshold still depends on its distance and sky position (Table~\ref{tab:xray}), so the eligible stars are those without measured evidence of activity above it. The nominal 2RXS depth reaches $\log L_X/L_{\rm bol}=-4$ at only about half of the AFGK stars, and Gaia does not flag every binary; the AFGK stars that these criteria miss remain eligible, so their natural emission would enter $k$ and weaken the limits rather than bias them low, and the post-cut tests show no excess. The envelope rests on the luminous 15\,GHz incoherent flare observed from DG~CVn and on the most luminous M-dwarf detections in this survey; a rarer flare above it would likewise be counted. The per-star limits are unaffected by eligibility: each bounds the total emission at the star's position. For the \NEmitters\ stars with emission (the \NNatural\ natural-emitter secure associations and \NCandStars\ forced detections, \NEmitPct\% of the sample), an artificial component is excluded only where it would exceed the measured emission: their limits are upper confidence bounds on the total flux density (Section~\ref{sec:limits_method}). Separating an artificial component from natural emission below $L_{\rm env}$ would require a model of each star's expected radio emission, for example from its X-ray luminosity through the G\"udel--Benz relation \citep{GudelBenz1993}, or from its variability and circular polarization, together with a search for an excess over it; we leave this to future work.

\subsection{Prospects} \label{sec:prospects}

VLASS~4.1 is the final survey campaign: it re-observed the half of the sky first covered in epoch 1.1, and its Single Epoch products will replace those from 1.1 \citep{Kimball2026}. The other half of the sky, including LSPM~J1943+5337, has three epochs. A targeted VLA observation of that source would extend its proper-motion baseline to nearly a decade and decide its nature, and similar observations, or any future extension of VLASS, would test every counted forced detection for co-motion. Single Epoch imaging improves the flux scale and astrometry, and its Stokes~V images would identify coherent stellar emission directly, as the Sydney Radio Star Catalogue does at lower frequencies \citep{Driessen2024}. Our limits average the received power over the full band, which is the right statistic for emission that fills it; a narrower signal is diluted. The planned VLASS Single Epoch coarse cubes, with one plane per 128\,MHz spectral window \citep{Kimball2026}, would improve the sensitivity to a signal confined to one spectral window by up to $\sqrt{16}=4$, and re-imaging the calibrated visibilities at the native 2\,MHz channel width, the filter width considered by \citet{Zuckerman2026}, by up to $\sim\sqrt{1000}\approx30$ for a signal within a single channel. VLBI observations would measure the brightness temperature and compactness that \citet{Garrett2026} proposes as the next stage of a broadband search. Lower-frequency continuum surveys extend the method to where coherent stellar emission and many terrestrial transmitters are strongest \citep{Callingham2021,Yiu2024}.

\section{Conclusions} \label{sec:conclusions}

We searched \NSample\ Gaia DR3 main-sequence stars within 100\,pc for excess 2--4\,GHz continuum emission in all VLASS epochs observed to date. Of \NSecure\ secure Epoch-1 associations, \NNatural\ are known natural emitters that, wherever the proper-motion baseline allows, move with their stars; the radio source near LSPM~J1943+5337 is not coincident with the star and is likely a background source. Forced photometry of \NWithForced\ stars in all epochs finds \NCandStars\ forced detections beyond the natural emitters. Applying the stellar-class and luminosity criteria---counting detections at apparently single AFGK stars outside known radio-emitter classes at any luminosity, and at M dwarfs and AFGK multiples only above $L_{\rm env}=10^{10}\,\WHz$---leaves \KSteady\ steady-test detection and \KHost\ stars detected in at least one epoch, consistent with the control positions; AFGK stars with Gaia multiplicity evidence show a significant excess of detections, as expected from active companions. Independent per-epoch catalogs, including Single Epoch and median-stack catalogs, confirm these results. We find no convincing technosignature candidates: the counted detections are no more numerous than chance coincidences, each is plausibly noise, a background source, or stellar emission, and none would warrant priority follow-up. We place limits on steady broadband emission at the position of every star with forced photometry (95\%-completeness thresholds, or upper confidence bounds where emission is detected)---median $\Lnu<\LlimAllMed\,\WHz$, and $\Lnu<\BarnardLAll\,\WHz$ for Barnard's star---and, with no background subtracted, limit the fraction of eligible AFGK stars that host a transmitter to $f<\FBzNine$ above $10^9\,\WHz$ ($\FBzWithMult$ without the multiplicity criterion), and of all stars outside the known-emitter classes to $f<\FBzFloor$ above $\sim\FloorL\,\WHz$. Counting every M-dwarf detection as a possible transmitter, without any eligibility decision, fewer than $\FMallNine$ of M dwarfs host one above $10^9\,\WHz$. Stacks detect population-averaged emission from M dwarfs, concentrated in those with activity evidence, and, for AFGK transmitters of a common luminosity, reach $f<1$ at luminosities \ReachRatioAFGK\ times fainter than do the individual detections, although only around the nearest stars. Continuum surveys are a modulation-agnostic complement to narrowband technosignature searches, covering signals that current narrowband pipelines do not search for, for every star in a nearly complete sample of nearby main-sequence stars.

\section{Data availability} \label{sec:data_avail}

All code, the downloaded catalogs, the \NCutoutsAll\ VLASS cutouts, and the derived tables and figures are archived at Zenodo, \dataset[doi:10.5281/zenodo.24326437]{https://doi.org/10.5281/zenodo.24326437}. Each script records the exact query used, and the package reproduces every number in this paper.

\begin{acknowledgments}

S.C. is supported by funding from Breakthrough Listen. The Breakthrough Prize Foundation funds the Breakthrough Initiatives, which manages Breakthrough Listen.

The National Radio Astronomy Observatory is a facility of the National Science Foundation operated under cooperative agreement by Associated Universities, Inc. CIRADA is funded by a grant from the Canada Foundation for Innovation 2017 Innovation Fund (Project 35999), as well as by the Provinces of Ontario, British Columbia, Alberta, Manitoba and Quebec. This research used the facilities of the Canadian Astronomy Data Centre operated by the National Research Council of Canada with the support of the Canadian Space Agency. This work has made use of data from the European Space Agency (ESA) mission Gaia, processed by the Gaia Data Processing and Analysis Consortium (DPAC). Funding for the DPAC has been provided by national institutions, in particular the institutions participating in the Gaia Multilateral Agreement. This research has made use of the SIMBAD database and the VizieR catalog access tool, CDS, Strasbourg, France, accessed via the CfA mirrors. The Digitized Sky Surveys were produced at the Space Telescope Science Institute under U.S. Government grant NAG W-2166. The Pan-STARRS1 Surveys have been made possible through contributions by the institutions listed in \citet{Chambers2016}. This work uses data from eROSITA, the soft X-ray instrument on the Russian--German SRG mission, processed with the eSASS software of the German eROSITA consortium, and from the ROSAT All-Sky Survey, retrieved through HEASARC (NASA Goddard Space Flight Center) and VizieR.

Analysis code and the manuscript text were prepared with assistance from an AI system (Claude Science / Opus 5.5, Anthropic). OpenAI GPT-6.2 Astra was also used to review the manuscript and code and suggest changes.  All results were verified by the author.
\end{acknowledgments}

\facilities{VLA, Gaia, ROSAT, eROSITA, PS1, PO:1.2m, MAST}

\software{Astropy \citep{Astropy2013,Astropy2018,Astropy2022}, NumPy \citep{Harris2020}, SciPy \citep{Virtanen2020}, pandas \citep{McKinney2010}, Matplotlib \citep{Hunter2007}, PyVO \citep{Graham2014}, PyBDSF \citep{Mohan2015}}

\clearpage
\bibliography{refs}{}
\bibliographystyle{aasjournalv7.1_patched}

\end{document}